\documentclass[prd,preprint,showpacs,preprintnumbers,amsmath,amssymb,eqsecnum,nofootinbib,groupedaddress,superscriptaddress,pdflatex]{revtex4-2}

\usepackage[pdftex]{graphicx}
\usepackage{bm,latexsym,amsmath,amssymb,amsfonts,mathrsfs}
\usepackage{float}
\usepackage{xcolor}
\usepackage{physics}
\usepackage{color}
\usepackage{comment}
\usepackage{here}
\usepackage{ulem}
\usepackage{empheq}

\newcommand{\PA}{{\partial}}

\allowdisplaybreaks[1]
\usepackage[pdftex,colorlinks=true,linkcolor=blue,citecolor=cyan,backref=page]{hyperref}
\makeatletter
\let\MYcaption\@makecaption
\makeatother

\usepackage{subcaption}
\makeatletter
\let\@makecaption\MYcaption
\makeatother
\makeatletter
\@addtoreset{equation}{section}

\makeatother
\begin{document}
\title{Cosmological long-wavelength solutions in non-adiabatic multi-fluid systems}
\author{Hayami~Iizuka}
\email[Email: ]{h.iizuka@rikkyo.ac.jp}
\affiliation{Department of Physics, Rikkyo University, Toshima, Tokyo 171-8501, Japan}
\author{Tomohiro~Harada}
\email[Email: ]{harada@rikkyo.ac.jp}
\affiliation{Department of Physics, Rikkyo University, Toshima, Tokyo 171-8501, Japan}

\date{\today}
%
\begin{abstract}
We develop a formulation of nonlinear cosmological perturbations on superhorizon scales in multi-fluid systems. It is based on the Arnowitt–Deser–Misner formalism combined with a spatial gradient expansion, characterized by a small parameter $\epsilon \equiv k/(a^{(b)}H)$, where $a(t)$ is the scale factor of the flat Friedmann-Lemaître-Robertson-Walker spacetime, ${}^{(b)}H(t)$ is the corresponding Hubble parameter, and $k$ is the comoving wavenumber.
Within this framework, we explicitly construct nonlinear long-wavelength solutions for cosmological perturbations. Since multi-fluid systems are inherently non-adiabatic, these solutions admit both adiabatic and entropy modes even at nonlinear leading order in the expansion. We define adiabatic and entropy perturbations, while face non uniqueness in defining pure entropy perturbations. 
Using the possible variations of pure entropy initial conditions, we analyze the time evolution of physical quantities, such as the curvature perturbation and density perturbation, in the geodesic slice for two-fluid systems.
\end{abstract}
\preprint{RUP-25-28}
\maketitle
\newpage

 \tableofcontents

\section{Introduction}
Primordial black holes (PBHs) are hypothetical black holes that could have formed in the early universe~\cite{Zeldovich:1967lct, Hawking:1971ei, Carr:1974nx}, particularly during the high-temperature, high-density epoch following the Big Bang. Unlike stellar black holes, which result from the gravitational collapse of massive stars, PBHs may have originated from the direct collapse of a large amplitude of density fluctuation generated in the early universe. As a result, their masses are not constrained by stellar evolution, and they can span a wide range. The PBH mass spectrum thus provides insight into small-scale cosmological perturbations that are far below the scales probed by the cosmic microwave background (CMB).

Various theoretical models have been proposed to explain the generation of curvature perturbations, which in turn generate density perturbations, particularly in the context of inflation and other early-universe scenarios. Given that the mass of PBHs is in the asteroid mass window, they are viable candidates for constituting the entirety of dark matter (see Ref.~\cite{Carr:2020gox} for a comprehensive review).
Recent advancements in gravitational wave (GW) observations, led by the LIGO-Virgo-KAGRA collaboration, have reported over a hundred events involving mergers of compact objects. PBHs are among the proposed sources of these events, offering a possible connection between PBHs and the observed GWs (see Ref.~\cite{Sasaki:2018dmp} for details on PBHs in the context of GWs).

Over the past three decades, cosmological nonlinear perturbation theory has made significant progress. A standard approach to derive non-linear long-wavelength solutions is the gradient expansion method
\cite{Tomita:1967wkp, Salopek:1990jq, Deruelle:1994iz, Shibata:1999yda},
which assumes that the characteristic physical length scale $L$ of interest is much larger than the Hubble radius ${}^{(b)}H^{-1}=(\dot{a}/a)^{-1}$, where $a$ is the scale factor of the fiducial flat Friedmann‐Lemaître‐Robertson-Walker (FLRW) spacetime. In this context, we choose the physical scale as $L^{-1}=k/a$, where $k$ is some wave number. Then, we can define a fictitious parameter $\epsilon\equiv k/(a{}^{(b)}H)$. The Einstein and matter field equations are then expanded in powers of $\epsilon$, yielding solutions known as cosmological long-wavelength solutions. 

Since gravitational collapse leading to the formation of a black hole is a strong gravitational phenomenon, simply solving the Einstein equations under approximations such as linear perturbation is insufficient to discuss gravitational collapse; the full Einstein equations must be solved. However, since the Einstein equations are nonlinear second-order partial differential equations, we cannot solve them analytically in general. Therefore, we need to use numerical simulation to solve these equations. Using the long-wavelength solutions as an initial condition for a numerical simulation of PBH formation is one of the interesting applications \cite{Shibata:1999yda, Harada:2015yda,Yoo:2021fxs,deJong:2021bbo}.

On cosmological scales, the universe is remarkably homogeneous and isotropic, as evidenced by the near-uniformity of CMB temperature and density fluctuations~\cite{COBE:1992syq, Planck:2018jri}. At these scales, we observe that the matter content of the universe behaves as an adiabatic fluid. However, our knowledge of the universe at smaller scales remains limited. In particular, the early universe probably underwent several transitions, such as reheating and the QCD phase transition, during which the dominant energy component changed. These transitional epochs naturally gave rise to systems composed of multiple fluids with distinct thermodynamic properties. In such settings, non-adiabaticity can play a role in the transitional phase.

Non-adiabatic perturbations naturally arise in multi-fluid systems, in contrast to the purely adiabatic fluctuations dominant in effectively single fluid systems. Non-adiabatic perturbations include the so-called entropy (isocurvature) perturbation, which describes fluctuations in the relative energy densities of different fluid components without affecting the total curvature. Recent studies suggest that such entropy fluctuations can be significantly amplified on small scales. These enhanced perturbations are of considerable interest, as they may act as sources of stochastic gravitational wave backgrounds with potentially observable spectra~\cite{Domenech:2021and} or trigger gravitational collapse leading to the formation of PBHs~\cite{Passaglia:2021jla,Yoo:2021fxs,deJong:2021bbo}.

Significant attention has been directed toward long-wavelength solutions in cosmological perturbation theory, particularly in the context of single-fluid systems~\cite{Tomita:1967wkp,Lyth:2004gb, Tanaka:2006zp, Tanaka:2007gh, Harada:2015yda}, as well as multi-component systems where one component overwhelmingly dominates the total energy density~\cite{Yoo:2021fxs}. However, in a general multi-fluid system, each element can contribute comparably to the energy budget of the universe, and the amplitude of its respective density fluctuations can be of similar magnitude. In such cases, both adiabatic and non-adiabatic perturbations can influence the leading-order behavior of the long-wavelength solutions. The formation of PBHs may occur during these multi-component epochs in the early universe. To simulate PBH formation in such systems, it is essential to construct consistent long-wavelength solutions that account for the non-negligible contributions from the multiple fluids. These solutions serve as physical initial conditions necessary for numerical simulations that follow the nonlinear gravitational collapse.

On superhorizon scales, the long-wavelength approximation provides a powerful framework in which the spacetime is approximated by a collection of locally FLRW ``separate universes"~\cite{Wands:2000dp}. The $\delta\mathscr{N}$ formalism, originating from the linear argument of Sasaki and Stewart~\cite{Sasaki:1995aw} and its nonlinear extension by Lyth, Malik, and Sasaki~\cite{Lyth:2004gb}, relies on this picture: the curvature perturbation $\zeta$ on a uniform-total-density slice is identified with the perturbation of the local e-folding number $\delta\mathscr{N}$. 
This identification has become one of the central tools in early-universe cosmology, particularly in studies of primordial non-Gaussianity in multi-component scenarios, including the curvaton model~\cite{Lyth:2005fi, Bartolo:2003jx}.
(See also Ref.~\cite{Sasaki:1998ug} for the superhorizon-scale dynamics in multi-scalar systems and its relation to the separate-universe picture.)
In an adiabatic system, once $\delta\mathscr{N}$ is specified as a function of the local state, the superhorizon evolution of scalar perturbations is determined by a single variable.
This property is automatically satisfied in single-field or effectively single-field systems, where the equation of state is barotropic, and the system is adiabatic at leading order in the gradient expansion.
In contrast, in non-adiabatic systems, Naruko, Takamizu, and Sasaki~\cite{Naruko:2012fe} investigated long-wavelength solutions in multi-scalar-field systems. 
They developed the so-called beyond-$\delta\mathscr{N}$ formalism, where the spatial dependence of the total energy density induces time dependence of the curvature perturbation already at $\mathcal{O}(\epsilon^0)$.

In multi-fluid systems with $N$ components, even if each component satisfies an independent conservation equation, we have $ N$ independent spatial integration functions.
These functions cannot be removed by gauge transformations and represent genuine physical degrees of freedom arising from intrinsic non-adiabaticity.
As a result, the spatial freedom that appears at $\mathcal{O}(\epsilon^0)$ is physical rather than gauge-related.
According to Kodama and Sasaki~\cite{Kodama:1984ziu, Kodama:1986fg, Kodama:1986ud}, Mukhanov et al.~\cite{Mukhanov:1990me}, Wands et al.~\cite{Wands:2000dp}, Lyth and Wands~\cite{Lyth:2003im} and Malik and Wands~\cite{Malik:2004tf}, one can construct conserved quantities even in the presence of non-adiabatic perturbations by making use of local energy conservation for each component.

In this paper, we construct explicit nonlinear long-wavelength solutions for general multi-fluid systems using the Arnowitt--Deser--Misner (ADM) formalism~\cite{Arnowitt:1959ah}. 
We provide a fully nonlinear definition of adiabatic and entropy perturbations applicable to multi-fluid cosmologies and present how to isolate an adiabatic perturbation and to define a pure entropy perturbation. We then apply this framework to two-fluid systems and explicitly demonstrate how curvature and density perturbations evolve when intrinsic non-adiabaticity contributes at leading order. 
Thus, building on the conventional $\delta \mathscr{N}$ picture, we 
establish a consistent theoretical foundation for the study of nonlinear perturbations in non-adiabatic multi-component universes.

This paper is organized as follows. We write down basic equations in the ADM formalism of the Einstein equation and perform the gradient expansion to obtain long-wavelength solutions in Sec.~\ref{sec2}. The definition of an adiabatic perturbation and an entropy perturbation in a multi-fluid system and the derivation of the long wavelength solution at zeroth-order are discussed in Sec.~\ref{sec3}.  
In Sec.~\ref{sec4}, we restrict our discussion to two-fluid systems and present the physical properties of the long-wavelength solutions.
As an example for the leading-order long-wavelength solutions expressed by elemental functions, we consider a matter-radiation system in Sec.~\ref{sec5}. Sec.~\ref{sec6} is devoted to concluding this paper.  In this paper, we use units in which both the speed of light and Newton’s gravitational constant are set to unity, $c=1~\text{and}~G=1$.
\section{Long-wavelength scheme in multi-fluid systems}
\label{sec2}
\subsection{Basic equations}
\subsubsection{ADM formalism}
We employ the ADM formalism~\cite{Arnowitt:1959ah}. The line element in four-dimensional spacetime has the following form:
\begin{equation}
    \dd s^2=-\alpha^2\dd t^2+\gamma_{ij}\qty(\beta^i\dd t+\dd x^i)\qty(\beta^j\dd t+\dd x^j),
\end{equation}
where $\alpha$, $\gamma_{ij}$, and $\beta^i$ are the lapse function, the induced metric, and the shift vector, respectively, and the Latin uppercase indices take values from 1 to 3. We lower these indices using $\gamma_{ij}$ and raise them using its inverse $\gamma^{ij}$. 

Let $h_{\mu\nu}$ be a projection tensor that projects four-dimensional physical quantities to the three-dimensional spacelike hypersurface $\Sigma_t$ as
\begin{equation}
    h_{\mu\nu}\equiv g_{\mu\nu}+n_\mu n_\nu,
\end{equation}
where $g_{\mu\nu}$ and $n_\mu$ are the four-dimensional metric and the normal one-form to $\Sigma_t$, respectively, with the normalisation condition as
\begin{equation}
    g^{\mu\nu}n_\mu n_\nu=-1.
\end{equation}
Thus, the covariant and contravariant components of the unit normal one-form and vector are given by
\begin{equation}
    n_\mu=\qty(-\alpha,0,0,0),\qq{and}~n^\mu=\qty(\frac{1}{\alpha},-\frac{\beta^i}{\alpha}),
\end{equation}
respectively.

The energy-momentum tensor for the matter field $T_{\mu\nu}$ is decomposed into
\begin{equation}
    T_{\mu\nu}=En_\mu n_\nu+J_\mu n_\nu + J_\nu n_\mu +S_{\rho\sigma}h^\rho_\mu h^\sigma_\nu,
\end{equation}
where $E$, $J_\mu$, and $S_{\mu\nu}$ are written as
\begin{equation}
    E\equiv T_{\mu\nu} n^\mu n^\nu,~J_\mu\equiv -T_{\sigma\nu} h^\sigma_\mu n^\nu \qq{and}~S_{\mu\nu}\equiv T_{\rho\sigma}h^\rho_\mu h^\sigma_\nu.
\end{equation}

The Einstein equation $G_{\mu\nu}=8\pi T_{\mu\nu}$ is decomposed 
into the Hamiltonian constraint $G^{\mu\nu}n_\mu n_\nu=8\pi T^{\mu\nu}n_\mu n_\nu$, 
the momentum constraint $G_{\mu\nu}n^\mu h_i^\nu=8\pi T_{\mu\nu}n^\mu h_i^\nu$, and the evolution equation
$G_{\mu\nu}h_i^\mu h_j^\nu=8\pi T_{\mu\nu}h_i^\mu h_j^\nu$.
These equations reduce to
\begin{align}
\mathcal{R}^k_k-K_{ij}K^{ij}+K^2=&~16\pi E,\label{hamiltonianconstraint}\\
\mathcal{D}_iK_j^i-\mathcal{D}_jK=&~8\pi J_j,\label{momentumconstraint}\\
\left(\partial_t-\beta^k\partial_k\right)K_{ij}=&~\alpha \mathcal{R}_{ij}-8\pi \alpha\left[S_{ij}+\frac{1}{2}\gamma_{ij}\left(E-S^k_k\right)\right]+\alpha \left(K_{ij}K-2K_{ik}K^{k}_j\right)\notag\\
  &-\mathcal{D}_j\mathcal{D}_i\alpha+K_{ik}\partial_j\beta^k+K_{jk}\partial_i\beta^k,\label{extirinsicevolution}
\end{align}
where $K_{ij}$, $\mathcal{R}_{ij}$, $\mathcal{R}$, and $\mathcal{D}_i$ are the extrinsic curvature, the Ricci tensor, the Ricci scalar, and the covariant derivative on the hypersurface $\Sigma_t$ with respect to $\gamma_{ij}$, respectively. The definition of the extrinsic curvature and the trace of extrinsic curvature $K$ is
\begin{equation}
    K_{ij}\equiv -h^\mu_i h^\nu_i\nabla_\mu n_\nu=-\frac{1}{2\alpha}\qty(\partial_t\gamma_{ij}-\mathcal{D}_i\beta_j-\mathcal{D}_j\beta_i),
\end{equation}
and 
\begin{equation}
    K\equiv\gamma^{ij}K_{ij}. 
\end{equation}

Let us consider a system with $N$ components of perfect fluids. An energy-momentum tensor for this system is given by
\begin{align}
  T_{\mu\nu}=&\sum_{\alpha=1}^{N}T_{(\alpha)\mu\nu},\\
  T_{(\alpha)\mu\nu}=&\qty(\rho_{(\alpha)}+p_{(\alpha)})u_{(\alpha)\mu} u_{(\alpha)\nu}+p_{(\alpha)}g_{\mu\nu},
  \end{align}
  where $\alpha=1,\dots, N$ are labels for each fluid, $\rho_{(\alpha)}~\text{and}~p_{(\alpha)}$ are energy density and pressure for each fluid, respectively.  The total energy-momentum tensor is decomposed as
  \begin{align}
    E&=\sum_{\alpha=1}^{N}\qty[\qty(\rho_{(\alpha)}+p_{(\alpha)})w_{(\alpha)}^2-p_{(\alpha)}],\label{ADMenergy}\\
    J_i&=\sum_{\alpha=1}^{N}\qty[\frac{w_{(\alpha)}^2}{\alpha}\qty(\rho_{(\alpha)}+p_{(\alpha)})\qty(v_{(\alpha)i}+\beta_i)],\label{ADMmomentum}\\
    S_{ij}&=\sum_{\alpha=1}^{N}\qty[p_{(\alpha)}\gamma_{ij}+\frac{J_{(\alpha)i}J_{(\alpha)j}}{E_{(\alpha)}+p_{(\alpha)}}],\label{ADMSij}
  \end{align}
  where $w_{(\alpha)}=\alpha u_{(\alpha)}^0$ and $v^i_{(\alpha)}=u^i_{(\alpha)}/u^0_{(\alpha)}$, respectively. We impose the normalization condition for each four-velocity: $g_{\mu\nu}u^\mu_{(\alpha)}u^\nu_{(\alpha)}=-1$.

  The total energy density and pressure are given by
  \begin{align}
    \rho_\text{tot}=&\sum_{\alpha=1}^{N}\rho_{(\alpha)},\\
    p_\text{tot}=&\sum_{\alpha=1}^{N}p_{(\alpha)}.
  \end{align} 

  We assume that fluid components are collisionless, which implies the following relation: 
  \begin{equation}
    \nabla_\mu T^{\mu\nu}_{(\alpha)}=0,\qq{for}\alpha=1,\dots,N.
  \end{equation}

  \subsubsection{Cosmological conformal decomposition}
We decompose the induced metric $\gamma_{ij}$ into the following form~\cite{Shibata:1999yda, Harada:2015yda}: 
 \begin{equation}
     \gamma_{ij}=a^2(t)\psi^4\tilde{\gamma}_{ij}.
 \end{equation}
 $\tilde{\gamma}_{ij}$ is chosen so that its determinant $\tilde{\gamma}$ is corresponding to that of the three-dimensional flat metric $\eta_{ij}$, which is independent of time. The function $a(t)$ is the scale factor of a fiducial flat FLRW spacetime. The extrinsic curvature is decomposed into
 \begin{equation}
     K_{ij}=\frac{\gamma_{ij}}{3}K + A_{ij},
 \end{equation}
 where $A_{ij}$ is the traceless part of the extrinsic curvature. We define the conformal decomposed traceless part of the extrinsic curvature $\tilde{A}_{ij}$ as follows:
 \begin{equation}
     A_{ij}\equiv a^2\psi^4\tilde{A}_{ij}\qq{or} A^{ij}\equiv a^{-2}\psi^{-4}\tilde{A}^{ij}.
 \end{equation}
 This shows $\tilde{\gamma}_{ij}\tilde{A}^{ij}=0$ by its definition.
 We raise and drop the Latin lowercase indices $i, j, k,\dots$ of tilded quantities by $\tilde{\gamma}^{ij}$ and $\tilde{\gamma}_{ij}$, respectively. We define the $\tilde{\mathcal{D}}_i$ and $\bar{\mathcal{D}}_i$ as the covariant derivatives with respect to $\tilde{\gamma}_{ij}$ and $\eta_{ij}$, respectively.
 We will denote the Laplacian with respect to $\tilde{\gamma}_{ij}$ and $\eta_{ij}$ as $\tilde{\Delta}\equiv\tilde{\gamma}^{ij}\tilde{\mathcal{D}}_i\tilde{\mathcal{D}}_j$ and $\bar{\Delta}\equiv\eta^{ij}\bar{\mathcal{D}}_i\bar{\mathcal{D}}_j$, respectively.

Using this decomposition, we find that the Hamiltonian and the momentum constraints are expressed by
\begin{gather}
  \tilde{\Delta}\psi= \frac{\tilde{\mathcal{R}}^k_k}{8}\psi-2\pi \psi^5a^2E-\frac{\psi^5a^2}{8}\qty(\tilde{A}_{ij}\tilde{A}^{ij}-\frac{2}{3}K^2),\label{hamiltonianconst}\\
  \tilde{\mathcal{D}}^j\qty(\psi^6\tilde{A}_{ij})-\frac{2}{3}\psi^6\tilde{\mathcal{D}}_iK=8\pi J_i\psi^6,\label{momentumconst}
\end{gather}
where $\tilde{\mathcal{R}}_{ij}$ and $\tilde{\mathcal{R}}^k_k$ are a Ricci tensor and Ricci scalar  with respect to $\tilde{\gamma}_{ij}$, respectively. It should be noted that the Ricci tensor is decomposed into the following form: 
\begin{equation}
\mathcal{R}_{ij}=\mathcal{R}^\psi_{ij}+\tilde{\mathcal{R}}_{ij},
\end{equation}
where 
\begin{align}
  \mathcal{R}^\psi_{ij}&\coloneqq-\frac{2}{\psi}\tilde{\mathcal{D}}_i\tilde{\mathcal{D}}_j\psi-\frac{2}{\psi}\tilde{\gamma}_{ij}\tilde{\Delta}\psi+\frac{6}{\psi^2}\tilde{\mathcal{D}}_i\psi\tilde{\mathcal{D}}_j\psi-\frac{2}{\psi^2}\tilde{\gamma}_{ij}\tilde{\mathcal{D}}_k\psi\tilde{\mathcal{D}}^k\tilde{\gamma},\\
  \tilde{\mathcal{R}}_{ij}&\coloneqq\frac{1}{2}\qty[-\bar{\Delta}\tilde{\gamma}_{ij}+\bar{\mathcal{D}}_i\bar{\mathcal{D}}^k\tilde{\gamma}_{kj}+\bar{\mathcal{D}}_j\bar{\mathcal{D}}^k\tilde{\gamma}_{ik}+2\bar{\mathcal{D}}_k\qty(f^{kl}C_{lij})-2C^l_{~kj}C^k_{~il}],\\
  \mathcal{R}^\psi&\coloneqq\tilde{\gamma}^{ij}R^\psi_{ij}=-\frac{8}{\psi^5a^2}\tilde{\Delta}\psi,\\
  f^{kl}&\coloneqq\tilde{\gamma}^{kl}-\eta^{kl},\\
C^k_{~ij}&\coloneqq\frac{1}{2}\tilde{\gamma}^{kl}\qty(\bar{\mathcal{D}}_i\tilde{\gamma}_{jl}+\bar{\mathcal{D}}_j\tilde{\gamma}_{il}-\bar{\mathcal{D}}_l\tilde{\gamma}_{ij}).
\end{align}

We also find the evolution equations for dynamical variables to be in the following form:
\begin{align}
  \qty(\PA_t-\mathcal{L}_\beta)\tilde{A}_{ij}=&\frac{1}{a^2\psi^4}\qty[\alpha\qty(\mathcal{R}_{ij}-\frac{\gamma_{ij}}{3}\mathcal{R})-\qty(\mathcal{D}_i\mathcal{D}_j\alpha-\frac{\gamma_{ij}}{3}\mathcal{D}_k\mathcal{D}^k\alpha)]+\alpha\qty(K\tilde{A}_{ij}-2\tilde{A}_{ik}\tilde{A}^k_j)\notag\\
  &-\frac{2}{3}\qty(\bar{\mathcal{D}}_k\beta^k)\tilde{A}_{ij}-\frac{8\pi\alpha}{a^2\psi^4}\qty(S_{ij}-\frac{\gamma_{ij}}{3}S^k_k),\label{cosmologicalconformaltracelessextrinsic}\\
  \qty(\PA_t-\mathcal{L}_\beta)\psi=&-\frac{\dot{a}}{2a}\psi+\frac{\psi}{6}\qty(-\alpha K+\bar{\mathcal{D}}_k\beta^k),\label{cosmologicalconformalconformalfactor}\\
  \qty(\PA_t-\mathcal{L}_\beta)K=&\alpha\qty(\tilde{A}_{ij}\tilde{A}^{ij}+\frac{1}{3}K^2)-\mathcal{D}_k\mathcal{D}^k\alpha+4\pi\alpha\qty(E+S^k_k),\label{cosmologicalconformaltraceextrinsic}\\
  \qty(\PA_t-\mathcal{L}_\beta)\tilde{\gamma}_{ij}=&-2\alpha\tilde{A}_{ij}-\frac{2}{3}\tilde{\gamma}_{ij}\bar{\mathcal{D}}_k\beta^k,\label{cosmologicalconformalthreemetevolution}
\end{align}
where $\mathcal{L}_\beta$ is Lie derivative along with the shift vector.
  The energy momentum conservation is decomposed into $n_\nu\nabla_\mu T_{(\alpha)}^{\mu\nu}=0,~\text{and}~\gamma_{\nu j}\nabla_\mu T_{(\alpha)}^{\mu\nu}=0$.
  The following equations give their concrete forms:
\begin{align}
  &\PA_t\qty[\psi^6a^3\qty{\qty(\rho_{(\alpha)}+p_{(\alpha)})w_{(\alpha)}^2-p_{(\alpha)}}]+\frac{1}{\sqrt{\eta}}\PA_l\qty[\sqrt{\eta}\psi^6a^3\qty{\qty(\rho_{(\alpha)}+p_{(\alpha)})w_{(\alpha)}^2-p_{(\alpha)}}v_{(\alpha)}^l]\notag\\
  &=-\frac{1}{\sqrt{\eta}}\PA_l\qty[\sqrt{\eta}\psi^6a^3p_{(\alpha)}\qty(v_{(\alpha)}^l+\beta^l)]+\alpha\psi^6a^3p_{(\alpha)}K
  -\frac{1}{\alpha}\PA_l\alpha\psi^6a^3w_{(\alpha)}^2\qty(\rho_{(\alpha)}+p_{(\alpha)})\qty(v_{(\alpha)}^l+\beta^l)\notag\\
  &+\frac{1}{\alpha}\psi^{10}a^5w_{(\alpha)}^2\qty(\rho_{(\alpha)}+p_{(\alpha)})\qty(v_{(\alpha)}^l+\beta^l)\qty(v_{(\alpha)}^m+\beta^m)\qty(\tilde{A}_{lm}+\frac{\tilde{\gamma}_{lm}}{3}K),\label{cosmoconenergyconse}\\
  &\PA_t\qty[w_{(\alpha)}\psi^6a^3\qty(\rho_{(\alpha)}+p_{(\alpha)})u_{(\alpha)j}]+\frac{1}{\sqrt{\eta}}\PA_k\qty[\sqrt{\eta}w_{(\alpha)}\psi^6a^3\qty(\rho_{(\alpha)}+p_{(\alpha)})v_{(\alpha)}^ku_{(\alpha)j}]=-\alpha\psi^6a^3\PA_jp_{(\alpha)}\notag\\
  &+w_{(\alpha)}\psi^6a^3\qty(\rho_{(\alpha)}+p_{(\alpha)})\qty(-w_{(\alpha)}\PA_j\alpha+u_{{(\alpha)}k}\PA_j\beta^k-\frac{u_{{(\alpha)}k}u_{{(\alpha)}l}}{2u_{(\alpha)}^0}\PA_j\gamma^{kl}),\label{cosmoconmomemtum}
\end{align}
for $\alpha=1,\dots,N$.

\subsection{Expansion scheme}
The spacetime is assumed to be smooth on a scale larger than $k^{-1}$. 
This assumption leads to the physical smoothing scale $L \equiv a/k$, where $a$ is the scale factor on flat FLRW spacetime.
In the context of cosmology, it is natural to choose a certain scale as a reference,
which is typically taken as $H \equiv \dot{a}/a$,
and use it to introduce the parameter $\epsilon$ defined as $\epsilon \equiv H^{-1} /L = k/(aH)$. 
Gradient expansion is the process of expanding the basic equations in powers of $\epsilon$,
assuming that spatial derivatives are small compared to the comoving Hubble scale: $\epsilon \ll 1$, 
allowing one to obtain solutions to any order. 
This corresponds to focusing on waves on superhorizon scales. 

Based on the above considerations, we assume two conditions. First, there exists a reference worldline $x^k=x_0^k$ such that $\psi(t,x_0^k)=1$.
Second, in the long-wavelength limit $\epsilon\to0$, the spacetime becomes locally homogeneous, isotropic, and spatially flat
\begin{equation}
  \dd s^2=-\dd t^2+a^2\eta_{ij}\dd x^i\dd x^j,
\end{equation}
where $\eta_{ij}$ is the three-flat metric on general coordinates.
Hence, we additionally assume $\psi=\order{\epsilon^0}$, $\beta^i=\order{\epsilon}$ and $\PA_t\tilde{\gamma}_{ij}=\order{\epsilon}$. Eq.~(\ref{cosmologicalconformalthreemetevolution}) and the assumption
$\PA_t\tilde{\gamma}_{ij}=\order{\epsilon}$ leads $\tilde{A}_{ij}=\order{\epsilon}$.

In the following discussion, the subscript in the upper left will denote the order of the gradient expansion, that is, for example,${^{(1)}}\tilde{A}_{ij}$ is the first order quantity of $\tilde{A}_{ij}$.

Moreover, we assume that the four-velocity for the individual fluid in the zeroth order is hypersurface orthogonal 
\begin{equation}
  u_{(\alpha)}^0=\frac{1}{\alpha}+\order{\epsilon^2},
\end{equation}
\begin{equation}
  v^i_{(\alpha)}=\order{\epsilon}.
\end{equation}
By this assumption, the two fluids can be regarded as a single, non-adiabatic perfect fluid in $\order{\epsilon^0}$.
Then, a total energy conservation laws : $-n_\nu\nabla_\mu T^{\mu\nu}=0$ reduces to 
\begin{equation}
\dot{\rho}_\text{tot}=-3\alpha\tilde{H}\qty(\rho_\text{tot}+p_\text{tot})+\order{\epsilon^2},\label{totalenergyconserve}
\end{equation}
where $\tilde{H}$ is the ``local'' Hubble parameter, which is defined by
\begin{equation}
  \tilde{H}\equiv\frac{1}{3}\nabla_\mu n^\mu.\label{deflocalhubble}
\end{equation}
From this definition, we find the relation: 
\begin{equation}
    \tilde{H}=-\frac{1}{3}K.
\end{equation}
Eq.~(\ref{totalenergyconserve}) implies
\begin{equation}
  \frac{\dot{a}}{a}+2\frac{\dot{\psi}}{\psi}=-\frac{1}{3}\frac{\dot{\rho}_\text{tot}}{\rho_\text{tot}+p_\text{tot}}+\order{\epsilon^2}.\label{totalenergyconserve2}
\end{equation}
Eq.~(\ref{totalenergyconserve}) also implies that we can expand the lapse function as
\begin{equation}
  \alpha={}^{(0)}\alpha(t,x^i)+\order{\epsilon^2},\label{lapseexpansion}
\end{equation}
where
\begin{equation}
  ^{(0)}\alpha(t,x^i)\equiv -\frac{^{(0)}\dot{\rho}_{\text{tot}}}{3{}^{(0)}\tilde{H}\qty(^{(0)}\rho_{\text{tot}}+^{(0)}p_{\text{tot}})}.\label{order1lapse}
\end{equation}

Let us expand the basic equations in a gradient expansion manner. The leading-order basic equations under general slice and threading conditions are given by 
\begin{align}
  {}^{(0)}K^2=&24\pi {}^{(0)}\rho_\text{tot},\label{beyondlytho(1)friedmann3}\\
  6\frac{\dot{\psi}}{\psi}+\alpha {}^{(0)}K=&-3\frac{\dot{a}}{a},\label{beyondlytho(1)hubble3}\\
  {}^{(0)}\dot{K}=&{}^{(0)}\alpha\qty[\frac{1}{3} {}^{(0)}K^2+4\pi{}^{(0)}\qty(E+S^i_i)],\label{beyondlytho(1)const3}\\
\PA_t\qty(a^3\psi^6{}^{(0)}\rho_{(\alpha)})=&{}^{(0)}\alpha a^3\psi^6
{}^{(0)}p_{(\alpha)}
{}^{(0)}K,\label{beyondlytho(1)energy3}
\end{align}
where
\begin{align}
  {}^{(0)}\rho_\text{tot}=& \sum_{\alpha=1}^{N}{}^{(0)}\rho_{(\alpha)},\label{leadingtptale0}\\
{}^{(0)}S^i_{i}=&3\sum_{\alpha=1}^{N}
{}^{(0)}p_{(\alpha)}
.
\end{align}

\subsection{Background solution}
We define the background spacetime $x^k=x^k_0$, where $\psi(x^k_0,t)=1$. The spacetime solution is exactly a flat FLRW at this spatial point. Thus, the local Hubble parameter is expressed as
\begin{equation}
  {}^{(b)}\tilde{H}=\frac{\dot{a}}{a},
\end{equation}
by the definition of local Hubble parameter Eq.~(\ref{deflocalhubble}). The subscript (b) in the upper left denotes background quantities.
Then, the mean curvature is expressed by
\begin{equation}
    {}^{(b)}K=-3\frac{\dot{a}}{a}.
\end{equation}
Moreover, Eq.~(\ref{ADMenergy}) shows
\begin{equation}
  {}^{(b)}E=\sum_{\alpha=1}^{N}{}^{(b)}\rho_{(\alpha)}={}^{(b)}E(t).
\end{equation}

In the following, we assume a barotropic EOS for each fluid, i.e., $p_{(\alpha)}=p_{(\alpha)}(\rho_{(\alpha)})$ for $\alpha=1,\dots, N$. While each fluid behaves as an adiabatic fluid, the entire system behaves as a non-adiabatic fluid. Moreover, just for simplicity and concreteness, we assume that each fluid has the EOS as follows:
  \begin{equation}
\Gamma_{(\alpha)}-1=\frac{p_{(\alpha)}}{\rho_{(\alpha)}},
  \end{equation}
  for $\alpha=1,\dots,N$. Although the present formulation itself does not require this EOS, with this assumption
Eq.~(\ref{cosmoconenergyconse}) for the background spacetime can be integrated to give
\begin{equation}
{}^{(b)}\rho_{(\alpha)}=D_{(\alpha)}a^{-3\Gamma_{(\alpha)}},\label{backgroundenergy}
\end{equation}
where $D_{(\alpha)}$ are an integral constant.
It leads to the Friedmann equation becoming
\begin{equation}
  \qty(\frac{\dot{a}}{a})^2=\frac{8\pi}{3}\sum_{\alpha=1}^{N}D_{(\alpha)}a^{-3\Gamma_{(\alpha)}}.\label{Haliltonianconstraint40}
\end{equation}

\subsection{Slicing conditions}
\label{relationsslice}
\label{relationslice}

The condition of the uniform Hubble (UH) slice is expressed by
\begin{equation}
\tilde{H} = \frac{\dot{a}}{a}.
\end{equation}
The condition of the constant mean curvature (CMC) slice is expressed by
\begin{equation}
K = -3\frac{\dot{a}}{a}.
\end{equation}
These expressions arise due to the assumption that the background spacetime is a spatially flat FLRW universe, where the expansion is homogeneous and isotropic. In such a background, the Hubble parameter $\dot{a}/a$ fully characterizes the extrinsic curvature of constant-time hypersurfaces, and the trace of the extrinsic curvature $K$ is directly proportional to the Hubble parameter. Therefore, these slices are equivalent to each of them. Therefore, we will refer to these slices as the CMC slice.

Taking the CMC slicing condition, the momentum constraint (\ref{momentumconst}) shows
\begin{equation}
  J_i=\order{\epsilon^2}.
\end{equation}
From this condition, we see that the comoving (C) slicing condition, which is defined by 
\begin{equation}
    J_i=0,
\end{equation}
and CMC slice are equivalent to $\order{\epsilon}$.
We can also conclude that the CMC slice corresponds to the uniform density (UD) slice, $\rho_\text{tot}={}^{(b)}\rho_\text{tot}(t)$, to $\order{\epsilon}$ from Eqs.~(\ref{beyondlytho(1)friedmann3}) and (\ref{leadingtptale0}). 
Therefore, we can identify the lapse function under the CMC slice by using Eqs.~(\ref{lapseexpansion}) and (\ref{order1lapse})
\begin{align}
  \alpha=&\frac{A(t)}{{}^{(b)}\rho_\text{tot}(t)+p_\text{tot}(t,x^k)}+\order{\epsilon^2},\label{condi.forlapse}
\end{align}
where $A(t)$ can be any function that depends only on $t$ and makes the lapse function positively definite.

 In the case of non-adiabatic fluids, as discussed in Ref.~\cite{Lyth:2004gb}, the lapse function at zeroth order is reduced to
\begin{align}
    {}^{(0)}\alpha=1-\frac{\delta p(t,x^k)}{{}^{(b)}\rho_\text{tot}
    (t)+p_\text{tot}\qty(t,x^k)},\label{non-adiabaticpressurelapse}
\end{align}
by taking the function $A(t)$ as the following form:
\begin{equation}
    A(t)={}^{(b)}\rho_{\text{tot}}(t)+{}^{(b)}p_{\text{tot}}(t),
\end{equation}
and $\delta p$ is defined by
\begin{equation}
    \delta p(t,x^k)\equiv{}^{(0)}p_\text{tot}(t,x^k)-{}^{(b)}p_{\text{tot}}(t).
\end{equation}
This function is a non-linear, non-adiabatic pressure perturbation at the zeroth order.

A proper time $\tau$ is defined by
\begin{equation}
    \tau(t,x^k)=\int_{x^k=\text{const.}}\alpha(t,x^k)\dd t,
\end{equation}
evaluated along the constant world line.
On the synchronous (Sync) slice, which is defined by
\begin{equation}
  \alpha=\alpha(t),\label{synchronous}
\end{equation} 
the proper time simplifies to a function of coordinate time only:
\begin{equation}
\tau\qty(t)\equiv\int^t\alpha\qty(\tilde{t})\dd \tilde{t}.\label{proper}
\end{equation}
 In general slices where $\alpha$ depends on both time and space, the proper time $\tau$ is not a coordinate. However, under the Sync slice, $\tau$ can be used as a new time coordinate, since it is purely a function of $t$. We can then redefine the time coordinate via  $\tilde{\alpha}\dd t=\dd \tau$.
In the following Sec.~\ref{2D}, we proceed with the discussion under the special case of this slicing condition: geodesic slice $\alpha=1$.
From this, comparing Eqs.~(\ref{non-adiabaticpressurelapse}) and (\ref{synchronous}), it can be seen that the CMC, UD, and comoving slices do not coincide with the geodesic slice (and Sync slice) even at $\order{\epsilon^0}$. 
If we restrict our discussion to G, the local Hubble parameter and mean curvature are not spatially constant even at $\order{\epsilon^0}$ (see Sec.~\ref{2D}). 
Since the Hamiltonian constraint~(\ref{beyondlytho(1)friedmann3}) shows a relation between the local Hubble parameter and total energy density, the total energy density is also not spatially constant at $\order{\epsilon^0}$ under the geodesic slice. 
It should also be noted that on the geodesic slice, the time coordinate $t$ coincides with the proper time $\tau$, and they share the same physical meaning.
Therefore, in what follows, we simply use $t$ as the time coordinate under the geodesic slicing condition. 

Taking the spatial gradient of both side of the Eq.~(\ref{totalenergyconserve2}) in geodesic slice, we find the following relation: 
\begin{equation}
    \PA_i\qty(\frac{\dot{\psi}}{\psi})=-\frac{1}{3}\PA_i\qty(\frac{\dot{\rho}_\text{tot}}{\rho_\text{tot}+p_\text{tot}})+\order{\epsilon^3}.\label{con}
\end{equation}
Integrating the above equation, since the first term of the right-hand side of Eq.~(\ref{con}) is $\order{\epsilon}$, we get the following relation: 
\begin{equation}
    \frac{\dot{\psi}}{\psi}=\order{\epsilon^0}.\label{dynamicalpsi}
\end{equation}
We find that the conformal factor $\psi$ acquires a time-dependent perturbation at $\order{\epsilon^0}$ in the presence of non-adiabaticity on the geodesic slice. This behavior contrasts with the case of adiabatic fluids considered in Refs.\cite{Lyth:2004gb, Tanaka:2006zp, Tanaka:2007gh, Harada:2015yda}, where the conformal factor remains constant in time to $\order{\epsilon}$. In multi-fluid systems, however, the conformal factor exhibits time dependence already at leading order, due to the presence of non-adiabatic perturbations. This result is consistent with earlier studies such as Ref.~\cite{Wands:2000dp}, which also emphasized the dynamical nature of curvature in non-adiabatic systems.

\section{Long-wavelength solutions in multi-fluid systems}
\label{sec3}
\subsection{Adiabatic and entropy perturbations}
Here, we introduce adiabatic and entropy perturbations without imposing that they are solutions of the basic equations.
\subsubsection{Adiabatic perturbation}
Let us consider an adiabatic perturbation and entropy perturbation conditions on a given time slice. 

The background solution for the matter energy density is given by Eq.~(\ref{backgroundenergy}):
\begin{equation}
{}^{(b)}\rho_{(\alpha)}=D_{(\alpha)}a^{-3\Gamma_{(\alpha)}}.
\end{equation}
Introducing the e-folding number as $\mathscr{N}(t)=\ln a(t)$, the energy density is reduced to
\begin{equation}
{}^{(b)}\rho_{(\alpha)}=D_{(\alpha)}e^{-3\Gamma_{(\alpha)}\mathscr{N}(t)}.\label{backgroundefoldingenergy}
\end{equation}
If there exists a common spatially dependent time shift $\delta\mathscr{N}(t,x^k)$ such that, for all the components $(\alpha)$,
\begin{equation}
\rho_{(\alpha)}=D_{(\alpha)}\qty[e^{\mathscr{N}(t)+\delta\mathscr{N}(t,x^k)}]^{-3\Gamma_{(\alpha)}},\label{pureadiabatic}
\end{equation}
then the perturbation is called an adiabatic perturbation. Perturbations that do not satisfy this are called non-adiabatic perturbations.

There is another, apparently different, definition of an adiabatic perturbation: if there exists a common time shift $\Delta t(t,x^k)$ such that, for all $(\alpha)$ 
\begin{equation}
    \rho_{(\alpha)}={}^{(b)}\rho_{(\alpha)}(t+\Delta t(t,x^k)),\label{sadiabatic}
\end{equation}
then this perturbation is usually called an adiabatic perturbation. This is equivalent to the definition given in Eq.~(\ref{pureadiabatic}), provided that $a(t)$ is a monotonic function of $t$. The reason is that replacing $t$ by $t+\Delta t(t,x^k)$ in the background solution is equivalent to replacing $\mathscr{N}$ by
\begin{align}
    \mathscr{N}(t+\Delta t(t,x^k))&=\mathscr{N}(t)+\mathscr{N}(t+\Delta t(t,x^k))-\mathscr{N}(t)\notag\\
    &=\mathscr{N}(t)+\delta \mathscr{N}(t,x^k),
\end{align}
where we have defined
\begin{equation}
    \delta\mathscr{N}(t,x^k)\equiv \mathscr{N}(t+\Delta t(t,x^k))-\mathscr{N}(t).
\end{equation}
The relation between this nonlinear definition and the linear perturbative notion of adiabaticity is further discussed in Appendix~\ref{linearnonlinearentropy}.

\subsubsection{Energy density at an equality time}
The total background energy density and the total energy density of the adiabatic perturbation are written as
\begin{align}
{}^{(b)}\rho_\text{tot}&=\sum_{\alpha=1}^N D_{(\alpha)}e^{-3\Gamma_{(\alpha)}\mathscr{N}(t)}={}^{(b)}\rho_\text{tot}(\mathscr{N};D_{(1)},\dots,D_{(N)}),\\
\rho_\text{tot}&=\sum_{\alpha=1}^N D_{(\alpha)}e^{-3\Gamma_{(\alpha)}(\mathscr{N}(t)+\delta \mathscr{N}(t,x^k))}={}^{(b)}\rho_\text{tot}(\mathscr{N}(t)+\delta\mathscr{N}(t,x^k);D_{(1)},\dots,D_{(N)}).
\end{align}
For an $N$-fluid background, we define the $\alpha\beta$–equal time for any two distinct components $\alpha$ and $\beta$ as the time at which the following condition is 
satisfied:
\begin{equation}
    {}^{(b)}\rho_{(\alpha)}={}^{(b)}\rho_{(\beta)}.
\end{equation}
From Eqs.~(\ref{backgroundenergy}) and~(\ref{backgroundefoldingenergy}), we find the following relation
\begin{align}
    a_{\alpha\beta-\text{eq}}&=\qty(\frac{D_{(\beta)}}{D_{(\alpha)}})^{1/\qty[3(\Gamma_{(\beta)}-\Gamma_{(\alpha)})]},\\
    \mathscr{N}_{\alpha\beta-\text{eq}}&\equiv\ln a_{\alpha\beta-\text{eq}},\\
    {}^{(b)}\rho_{(\alpha)\alpha\beta-\text{eq}}&={}^{(b)}\rho_{(\beta)\alpha\beta-\text{eq}}=\qty(D_{(\alpha)}^{\Gamma_{(\beta)}}D_{(\beta)}^{-\Gamma_{(\alpha)}})^{1/(\Gamma_{(\beta)}-\Gamma_{(\alpha)})},\\
    {}^{(b)}\rho_{\text{tot},\alpha\beta-\text{eq}}&={}^{(b)}\rho_\text{tot}\qty( \mathscr{N}_{\alpha\beta-\text{eq}};D_{(1)},\dots,D_{(N)}).
\end{align}
For the adiabatic perturbation Eq.~(\ref{pureadiabatic}), we find the same relation for $\mathscr{N}+\delta\mathscr{N}$
\begin{align}
\qty[e^{\mathscr{N}+\delta\mathscr{N}}]_{\alpha\beta-\text{eq}}&=\qty(\frac{D_{(\beta)}}{D_{(\alpha)}})^{1/\qty[3(\Gamma_{(\beta)}-\Gamma_{(\alpha)})]},\\
    \rho_{(\alpha)\alpha\beta-\text{eq}}(x^k)&=\rho_{(\beta)\alpha\beta-\text{eq}}=\qty(D_{(\alpha)}^{\Gamma_{(\beta)}}D_{(\beta)}^{-\Gamma_{(\alpha)}})^{1/(\Gamma_{(\beta)}-\Gamma_{(\alpha)})},\\
    \rho_{\text{tot},\alpha\beta-\text{eq}}&=\rho_\text{tot}(\qty(\mathscr{N}+\delta\mathscr
    {N})_{\alpha\beta-\text{eq}};D_{(1)},\dots,D_{(N)})=\rho_\text{tot}\qty( \mathscr{N}_{\alpha\beta-\text{eq}};D_{(1)},\dots,D_{(N)}).
\end{align}
Therefore, the adiabatic energy density is the same as the background one at $\alpha\beta$-equality time.

\subsubsection{Entropy perturbation}
\label{entropy}
Solving the adiabatic condition, Eq.~(\ref{pureadiabatic}), for $\mathscr{N}+\delta\mathscr{N}$, we find the following solution
\begin{equation}
    \mathscr{N}(t)+\delta\mathscr{N}(t,x^k)=-\frac{1}{3\Gamma_{(\alpha)}}\ln\qty(\frac{\rho_{(\alpha)}}{D_{(\alpha)}}),\qq{or}\delta\mathscr{N}(t,x^k)=-\frac{1}{3\Gamma_{(\alpha)}}\qty(\frac{\rho_{(\alpha)}}{{}^{(b)}\rho_{(\alpha)}}).
\end{equation}
From this solution, we conclude that $-1/(3\Gamma_{(\alpha)})\left(\rho_{(\alpha)}/({}^{(b)}\rho_{(\alpha)})\right)$ does not depend on a component of $(\alpha)$ for the adiabatic perturbation. This is rewritten as
\begin{equation}
    -\frac{1}{3\Gamma_{(\alpha)}}\qty(\frac{\rho_{(\alpha)}}{{}^{(b)}\rho_{(\alpha)}})=-\frac{1}{3\Gamma_{(\alpha)}}\ln\qty(1+\delta_{(\alpha)}).
\end{equation}
Thus, the necessary and sufficient condition for the existence of the function $\delta\mathscr{N}$ is for an arbitrary pair of components, $(\alpha),~(\beta)$, satisfying
\begin{equation}
    -\frac{1}{3\Gamma_{(\alpha)}}\ln\qty(1+\delta_{(\alpha)})=-\frac{1}{3\Gamma_{(\beta)}}\ln\qty(1+\delta_{(\beta)}).
\end{equation}
Then, it is reasonable to define an entropy perturbation $S_{(\alpha\beta)}$ as
\begin{equation}
    S_{(\alpha\beta)}\equiv \frac{1}{3\Gamma_{(\alpha)}}\ln\qty(1+\delta_{(\alpha)})-\frac{1}{3\Gamma_{(\beta)}}\ln\qty(1+\delta_{(\beta)}).\label{entropyperturbation}
\end{equation}

\subsection{Zeroth-order adiabatic and entropy perturbations}
Here, we construct zeroth-order long wavelength solutions and decompose them into adiabatic and entropy perturbations.
\subsubsection{Zeroth-order solutions}
Integrating Eqs.~(\ref{beyondlytho(1)hubble3}) and (\ref{beyondlytho(1)energy3}), we find the energy density at the zeroth order under the general slice:
\begin{equation}
    {}^{(0)}\rho_{(\alpha)}(t,x^k)=C_{(\alpha)}(x^k)\qty(a(t)\psi^2)^{-3\Gamma_{(\alpha)}},
\end{equation}
where $C_{(\alpha)}(x^k)$ is an integral function which comes from integration of the energy equation for each fluid. 
For later convenience, we decompose the integral function as follows:
\begin{equation}
    \bar{C}_{(\alpha)}(x^k)\equiv \frac{C_{(\alpha)}(x^k)}{D_{(\alpha)}}.\label{decompositionofintegralfunc.}
\end{equation}
Then, the energy density for each fluid is rewritten by
\begin{equation}
{}^{(0)}\rho_{(\alpha)}=D_{(\alpha)}\bar{C}_{(\alpha)}(x^k)(a(t)\psi^2)^{-3\Gamma_{(\alpha)}}.\label{3.35}
\end{equation}

Following the convention of Lyth, Malik, and Sasaki (2005)~\cite{Lyth:2004gb}, we define a local e-folding number $\mathscr{N}(t_2,t_1;x^k)$ as follows:
\begin{equation}
    \mathscr{N}(t_2,t_1;x^k)\equiv-\frac{1}{3}\int^{t_2}_{t_1,x^k=\text{const}}\alpha K\dd t.
\end{equation}
Using Eq.~(\ref{beyondlytho(1)hubble3}), this definition is reduced to
\begin{equation}
    \mathscr{N}(t_2,t_1;x^k)=\ln \frac{a(t_2)}{a(t_1)}+2\ln\frac{\psi(t_2,x^k)}{\psi(t_1,x^k)}.
\end{equation}

For the background solution, since we take time slicing which coincides with the natural one of the FLRW solution with $\psi=1$, we do not need to take care of the spatial dependence of the function $\mathscr{N}$. Also, taking the lower end of the integral reference time $t_0$, the background e-folding number is defined by
\begin{equation}
    \mathscr{N}(t)\equiv\mathscr{N}(t,t_0;x^k), 
\end{equation}
where $t_{0}$ on the left-hand side is omitted for brevity.
If we redefine $D_{(\alpha)} a(t_0)^{-3\Gamma_{(\alpha)}}$ simply as $D_{(\alpha)}$, then all the equations obtained so far can be used without any further modification.

By choosing the lower limit also at $t_0$ for the zeroth–order solution, we can write
\begin{equation}
    {}^{(0)}\rho_{(\alpha)}(t,x)
    = \bar{C}_{(\alpha)}(x)\, D_{(\alpha)}
    e^{-3\Gamma_{(\alpha)}\, \mathscr{N}(t,t_0; x)} ,
    \label{eq:rho_zero_order}
\end{equation}
where we have redefined 
$\bar{C}_{(\alpha)}(x)\psi^{-6\Gamma_{(\alpha)}}(t_0,x)$ 
again as $\bar{C}_{(\alpha)}(x)$.
In what follows, we simply denote $\mathscr{N}(t,t_0; x)$ by $\mathscr{N}$ and treat it as if it were
a time coordinate.  
With this choice, the gauge dependence of $\psi(t,x)$ can be absorbed into $\mathscr{N}$.

\subsubsection{Adiabatic perturbation}
We now discuss the necessary and sufficient condition for the zeroth-order solution (\ref{eq:rho_zero_order}) to be an adiabatic perturbation. Substituting the zeroth-order solution Eq.~(\ref{eq:rho_zero_order}) into 
Eq.~(\ref{pureadiabatic}), we obtain
\begin{equation}
    \bar{C}_{(\alpha)}(x^k)D_{(\alpha)}e^{-3\Gamma_{(\alpha)}\mathscr{N}}=D_{(\alpha)}e^{-3\Gamma_{(\alpha)}\qty(\mathscr{N}+\delta\mathscr{N})}.
\end{equation}
Here, assuming that the values of $\mathscr{N}$ on both sides are identical, we find that the existence 
of a function $\delta\mathscr{N}^\text{ad}(x^k)$, independent of $(\alpha)$, such that the expression reduces to 
\begin{equation}
    -\frac{\ln\bar{C}_{(\alpha)}(x^k)}{3\Gamma_{(\alpha)}}=\delta\mathscr{N}^\text{ad}(x^k),\qq{or} \bar{C}_{(\alpha)}(x^k)=e^{-3\Gamma_{(\alpha)}\delta\mathscr{N}^\text{ad}(x^k)},\label{ASCforadi.}
\end{equation}
constitutes the necessary and sufficient condition for the perturbation to be adiabatic. From this, we also find the following relation:
\begin{equation}
    \delta \mathscr{N}=\delta\mathscr{N}^\text{ad}(x^k).
\end{equation}
The background solution can be obtained by setting $\bar{C}=1$ and $\psi=1$. 
Therefore, an adiabatic perturbation possesses one degree of freedom, corresponding to 
an arbitrary function of $x^k$. Assuming that the zeroth-order solution is an adiabatic perturbation, Eq.~(\ref{eq:rho_zero_order}) can be rewritten for all $ (\alpha)$ as
\begin{equation}
{}^{(0)}\rho_{(\alpha)}
=D_{(\alpha)}e^{-3\Gamma_{(\alpha)}\mathscr{N}}
e^{-3\Gamma_{(\alpha)}\delta\mathscr{N}^\mathrm{ad}}=D_{(\alpha)}e^{-3\Gamma_{(\alpha)}(\mathscr{N}+\delta\mathscr{N}^\mathrm{ad})}.
\label{eq:rho_adiabatic}
\end{equation}

Up to this point, the discussion has been carried out at a fixed time $t$, but in fact, 
the condition on $C_{(\alpha)}(x^k)$ using the actual zeroth-order solution, Eq.~(\ref{ASCforadi.}), 
is independent of $t$. Thus, it can be shown that a perturbation that is adiabatic at 
a given time is adiabatic at any other time. It should be noted that the temporal and spatial dependence of $\psi$ for an adiabatic perturbation is not fixed unless the gauge is specified.

By setting ${}^{(0)}K=-3\tilde{H}$, Eq.~(\ref{beyondlytho(1)friedmann3}) gives
\begin{equation}
    \tilde{H}^2 = \frac{8\pi}{3} {}^{(b)}\rho_\text{tot}.
\end{equation}
Therefore, in the two-fluid case, for a zeroth-order adiabatic perturbation, the expansion rate at the equal-density time is independent of $x^k$ and coincides with the background one. 
This result naturally extends to the multi-fluid case: for a zeroth-order adiabatic perturbation, the expansion rate at the $\alpha\beta$-equality time is also independent of $x^k$ and equal to the background one.

\subsubsection{Entropy perturbation}
\label{sec:defofentropy}
We introduce a function $\delta \mathscr{N}_{(\alpha)}(x^k)$ as
\begin{equation}
    \bar{C}_{(\alpha)}(x^k)=e^{-3\Gamma_{(\alpha)}\delta\mathscr{N}_{(\alpha)}(x^k)},
\end{equation}
for later convenience. Then, Eq.~(\ref{eq:rho_zero_order}) can be rewritten as
\begin{equation}
    {}^{(0)}\rho_{(\alpha)}=D_{(\alpha)}e^{-3\Gamma_{(\alpha)}\qty(\mathscr{N}+\delta \mathscr{N}_{(\alpha)}(x^k))},
\end{equation}
for all $(\alpha)$. In this way, each point $x^k$ can be interpreted as being described by a separate universe on top of a common background, in which the evolution of each species $(\alpha)$—whether faster 
or slower—is specified by $\delta \mathscr{N}_{(\alpha)}(x^k)$.

Since an adiabatic perturbation is represented by $\delta\mathscr{N}^\text{ad}(x^k)$, the quantity obtained by dividing Eq.~(\ref{eq:rho_zero_order}) by Eq.~(\ref{eq:rho_adiabatic}), namely
\begin{equation}
\frac{{}^{(0)}\rho_{(\alpha)}}{{}^{(0)}\rho_{(\alpha)}|_{\text{ad}}} =e^{-3\Gamma_{(\alpha)}\qty(\delta\mathscr{N}_{(\alpha)}(x^k)-\delta\mathscr{N}^\text{ad}(x^k))}=e^{-3\Gamma_{(\alpha)}\delta\mathscr{N}^\text{nad}_{(\alpha)}(x^k)},
    \label{eq:entropy_def}
\end{equation}
constitutes an independent degree of freedom, where
\begin{equation}
    \delta\mathscr{N}^\text{nad}_{(\alpha)}(x^k)\coloneqq \delta\mathscr{N}_{(\alpha)}(x^k)-\delta\mathscr{N}^\text{ad}(x^k).
\end{equation}
This degree of freedom exists regardless of the presence of an adiabatic perturbation and is therefore naturally referred to as a non-adiabatic perturbation. Since $\delta\mathscr{N}^\text{ad}(x^k)$ can be chosen arbitrarily, a non-adiabatic perturbation contains one arbitrary function of $x^k$ corresponding to the adiabatic perturbation, and $(N-1)$ arbitrary functions of $x^k$ corresponding to entropy perturbations. With this definition, it follows naturally that if an entropy perturbation does not exist at a given time, it does not exist at any other time.

For $N$ fluid systems, an entropy perturbation has $(N-1)$ degrees of freedom, for example, corresponding to $\{\delta\mathscr{N}_i-\delta\mathscr{N}_{i+1}\}_{i=1,\dots,N-1}$. Using Eq.~(\ref{eq:rho_zero_order}), we obtain
\begin{align}
    e^{\tilde{\mathscr{N}}_{\alpha\beta-\text{eq}}(x^k)}&=\qty(\frac{\bar{C}_{(\beta)}{D_{(\beta)}}}{\bar{C}_{(\alpha)}{D_{(\alpha)}}})^{1/\qty[3\qty(\Gamma_{(\beta)}-\Gamma_{(\alpha)})]},\\
    \tilde{\rho}_{(\alpha),\alpha\beta-\text{eq}}(x^k)&=\tilde{\rho}_{(\beta),\alpha\beta-\text{eq}}(x^k)=\qty[\qty(\bar{C}_{(\alpha)}{D_{(\alpha)}})^{\Gamma_{(\beta)}}\qty(\bar{C}_{(\beta)}{D_{(\beta)}})^{-\Gamma_{(\alpha)}}]^{1/\qty(\Gamma_{(\beta)}-\Gamma_{(\alpha)})}.
\end{align}
It then follows that
\begin{equation}
    h_{(\alpha\beta)}^2(x^k) \equiv 
    \frac{\tilde{\rho}_{(\alpha),\alpha\beta-\text{eq}}(x^k)}{{}^{(b)}\rho_{(\alpha),\alpha\beta-\text{eq}}} 
    = \left(\frac{\bar{C}_{(\alpha)}(x)^{1/\Gamma_{(\alpha)}}}{\bar{C}_{(\beta)}(x)^{1/\Gamma_{(\beta)}}}\right)^{\frac{\Gamma_{(\beta)} \Gamma_{(\alpha)}}{\Gamma_{(\beta)}-\Gamma_{(\alpha)}}}
=e^{3\Gamma_{(\alpha)}\Gamma_{(\beta)}\qty(\delta\mathscr{N}_{(\alpha)}-\delta\mathscr{N}_{(\beta)})/\qty(\Gamma_{(\alpha)}-\Gamma_{(\beta)})},
\end{equation}
so that
\begin{equation}
    \delta\mathscr{N}_{(\alpha\beta)} \equiv \frac{1}{2} (\delta\mathscr{N}_{(\alpha)} - \delta\mathscr{N}_{(\beta)} )
    =\frac{1}{2}\qty(\delta\mathscr{N}^\text{nad}_{(\alpha)}-\delta\mathscr{N}^\text{nad}_{(\beta)})
    = \frac{\Gamma_{(\alpha)} - \Gamma_{(\beta)}}{3 \Gamma_{(\alpha)} \Gamma_{(\beta)}} \ln h_{(\alpha\beta)}.\label{def:entropy}
\end{equation}
This procedure allows us to isolate the $\alpha\beta$-entropy perturbation. In fact, the entropy perturbation $S_{(\alpha\beta)}$ defined in Sec.~\ref{entropy} is rewritten by
\begin{equation}
    S_{(\alpha\beta)}=-2\delta\mathscr{N}_{(\alpha\beta)}.
\end{equation}

Both $S_{(\alpha\beta)}$ and $\delta\mathscr{N}_{(\alpha\beta)}$ are fully non-linear 
definitions of the $\alpha\beta$-entropy perturbation. 
The quantity $\delta\mathscr{N}^\text{nad}_{(\alpha)}$ represents the non-adiabatic, non-linear 
degree of freedom of each fluid component individually. 
$\delta\mathscr{N}_{(\alpha\beta)}$ can be viewed as the difference $(\delta\mathscr{N}^\text{nad}_{(\alpha)}-\delta\mathscr{N}^\text{nad}_{(\beta)})$
that isolates the $\alpha\beta$-entropy perturbation. 
All three quantities describe the same physical degrees of freedom, 
but in different representations. 
In the linearized limit, they reduce to the conventional definition of the entropy (isocurvature) perturbation familiar in linear perturbation theory. Accordingly, for separately conserved barotropic components, the spatial integration functions $\bar{C}_{(\alpha)}(x^k)$ (or equivalently $\delta\mathscr{N}_{(\alpha)}$) encode the component-wise conserved quantities in the separate-universe picture.
In this sense, they are the non-linear counterparts to the conserved linear perturbations associated with each fluid discussed in Refs.~\cite{Kodama:1986fg, Kodama:1986ud, Kodama:1984ziu, Mukhanov:1990me, Lyth:2003im}, up to slicing and sign/normalization conventions.
Relative combinations such as $\delta\mathscr{N}_{(\alpha)}-\delta\mathscr{N}_{(\beta)}$ represent nonlinear entropy (isocurvature) degrees of freedom.

Let us restrict our discussion to a two-fluid system. 
We consider the expansion rate at the equal-density hypersurface. 
As we have already seen, for an adiabatic perturbation, the expansion rate at equality time is uniform and equal to the background value. 
For a non-adiabatic perturbation, using Eq.~(\ref{eq:rho_zero_order}), we find
\begin{align}
    e^{\tilde{\mathscr{N}}_\mathrm{eq}(x^k)} &= 
    \left( \frac{\bar{C}_{(2)}(x^k) D_{(2)}}{\bar{C}_{(1)}(x^k) D_{(1)}} \right)^{\frac{1}{3(\Gamma_{(2)} - \Gamma_{(1)})}}, \\
    \tilde{H}_\mathrm{eq}(x^k) &= \frac{8\pi}{3} \rho_{\rm tot, eq}(x^k) = 
    \frac{8\pi}{3} \left[ (\bar{C}_{(1)} D_{(1)})^{\Gamma_{(2)}} (\bar{C}_{(2)} D_{(2)})^{-\Gamma_{(1)}} \right]^{\frac{1}{\Gamma_{(2)} - \Gamma_{(1)}}}.
\end{align}
It then follows that
\begin{equation}
    h(x^k) \equiv \frac{\tilde{H}_\mathrm{eq}(x^k)}{H_\mathrm{eq}} 
    = \left( \frac{\bar{C}_{(1)}(x^k)^{1/\Gamma_{(1)}}}{\bar{C}_{(2)}(x^k)^{1/\Gamma_{(2)}}} \right)^{\frac{\Gamma_{(1)} \Gamma_{(2)}}{2\qty(\Gamma_{(1)} - \Gamma_{(2)})}} 
    = \exp\left[ \frac{3 \Gamma_{(1)} \Gamma_{(2)}}{2\qty(\Gamma_{(1)} - \Gamma_{(2)})} (\delta \mathscr{N}_{(1)} - \delta \mathscr{N}_{(2)}) \right].\label{dimlesshubble}
\end{equation}
Consequently, we can define the two-fluid entropy perturbation as
\begin{equation}
    \delta\mathscr{N}_{(12)} = \frac{1}{2} (\delta \mathscr{N}_{(1)} - \delta \mathscr{N}_{(2)}) 
    = \frac{\Gamma_{(1)} - \Gamma_{(2)}}{3 \Gamma_{(1)} \Gamma_{(2)}} \ln h(x^k),
\end{equation}
which successfully isolates the entropy perturbation between the two fluids.

\subsection{Defining pure entropy perturbations}
\label{isolation}
From the above discussion, we find that adiabatic perturbations possess an intrinsic arbitrariness.
In the following subsubsection, we fix this arbitrariness and attempt to isolate adiabatic perturbations by adopting a concrete profile for the adiabatic mode.
\subsubsection{Primeval isocurvature condition}
\label{iso:inflation}
A simple way to fix the choice of $\delta\mathscr{N}^\text{ad}(x^k)$ is, for example,
\begin{equation}
    \delta\mathscr{N}^\text{ad} = \delta \mathscr{N}_{(1)}, \quad \text{or equivalently} \quad 
    \bar{C}_{(1)} = e^{-3\Gamma_{(1)}\delta\mathscr{N}^\text{ad}}.\label{isolation:inflation}
\end{equation}
The remaining functions $\bar{C}_{(\alpha)}(x)$ for $\alpha = 2, \dots, N$ can then be chosen freely.
In this choice, the total density fluctuation ${}^{(0)}\delta_\text{tot}$, which is defined by
\begin{equation}
{}^{(0)}\delta_\text{tot}\equiv\frac{{}^{(0)}\rho_\text{tot}-{}^{(b)}\rho_\text{tot}}{{}^{(b)}\rho_\text{tot}}
    =\frac{\sum_{\alpha}{}^{(b)}\rho_{(\alpha)}{}^{(0)}\delta_{(\alpha)}}{\sum_\beta{}^{(b)}\rho_{(\beta)}},
\end{equation}
vanishes at the early universe, where the component $(1)$ dominates, but subsequently can grow nonlinearly due to the entropy perturbation.
In the two-fluid case, this choice leads to
\begin{align}
    \delta \mathscr{N}_{(1)} &= \delta\mathscr{N}^\text{ad}, \\
    \delta \mathscr{N}_{(2)} &= \delta\mathscr{N}^\text{ad} - 2 \delta\mathscr{N}_{(12)},
\end{align}
so that $\delta \mathscr{N}_{(1)}$ and $\delta \mathscr{N}_{(2)}$ are expressed as linear combinations of $\delta \mathscr{N}^\text{ad}$ and $\delta\mathscr{N}_{(12)}$.
With this convention, the pure entropy condition corresponds to setting $\delta\mathscr{N}^\text{ad} = 0$, which immediately implies $\delta \mathscr{N}_{(1)} = 0$. As we will see later in Sec.~\ref{initially isocurvature}, this choice corresponds to the initially isocurvature condition in two-fluid systems in geodesic slicing. 

\subsubsection{Zero-average condition}
\label{sym:isolation}
An another choice for $\delta\mathscr{N}^\text{ad}(x^k)$ is given by
\begin{equation}
    \delta\mathscr{N}^\text{ad} = \frac{1}{N} \sum_{\alpha} \delta \mathscr{N}_{(\alpha)}(x^k).\label{isolation:symmetric}
\end{equation}
This can be interpreted as the reference adiabatic perturbation corresponding to a separate-universe picture obtained by averaging the $\delta \mathscr{N}_{(\alpha)}$ over all components $(\alpha)$ in the original non-adiabatic perturbation. 

In the two-fluid case, this gives
\begin{equation}
    e^{\delta\mathscr{N}^\text{ad}(x^k)} = \left[ \bar{C}_{(1)}(x^k)^{1/\Gamma_{(1)}}\bar{C}_{(2)}(x^k)^{1/\Gamma_{(2)}} \right]^{-1/6}.\label{symmetricpriscription}
\end{equation}
Consequently, $\delta \mathscr{N}_{(1)} + \delta \mathscr{N}_{(2)}$ corresponds to the adiabatic perturbation, while $\delta \mathscr{N}_1 - \delta \mathscr{N}_{(2)}$ corresponds to the entropy perturbation. 
Furthermore, combining Eqs.~(\ref{symmetricpriscription}) and (\ref{dimlesshubble}), we can express $\delta \mathscr{N}_{(1)}$ and $\delta \mathscr{N}_{(2)}$ as linear combinations of $\delta\mathscr{N}^\text{ad}$ and $\delta\mathscr{N}_{(12)}$:
\begin{align}
    \delta \mathscr{N}_{(1)} &= \delta\mathscr{N}^\text{ad} + \delta\mathscr{N}_{(12)},\label{sym:two1} \\
    \delta \mathscr{N}_{(2)} &= \delta\mathscr{N}^\text{ad} - \delta\mathscr{N}_{(12)}.\label{sym:two2}
\end{align}
The pure entropy condition corresponds to setting $\delta\mathscr{N}^\text{ad} = 0$, which  implies $\delta \mathscr{N}_{(1)} = -\delta \mathscr{N}_{(2)}$.

\subsubsection{Initially isocurvature condition}
\label{iso:isolation}
Alternatively, one can use the freedom in choosing $\delta\mathscr{N}^\text{ad}$ to fix $\bar{C}_{(\alpha)}(x^k)$ such that, under a certain gauge condition, 
\begin{equation}
    \psi(t_0,x^k) = 1.
\end{equation}
Here, we consider one example of this approach.

The zeroth-order solution ${}^{(0)}\rho_{(\alpha)}(t,x^k)$ is integrated as in Eq.~(\ref{3.35}), where the background solution ${}^{(b)}\rho_{(\alpha)}(t)$ is multiplied by $\bar{C}_{(\alpha)}(x^k) \psi^{-6\Gamma_{(\alpha)}}$. The total density $\rho_\mathrm{tot}(t,x^k)$ can be computed from Eq.~(\ref{leadingtptale0}). Then, for the zeroth-order solution on the geodesic slice (taking the lapse function $\alpha=1$), Eqs.~(\ref{beyondlytho(1)friedmann3}) and (\ref{beyondlytho(1)hubble3}) imply that $a \psi^2$ exactly satisfies the Friedmann equation with the background $D_{(\alpha)}$ replaced by $\bar{C}_{(\alpha)} D_{(\alpha)}$. 

If one further imposes the no-decaying-mode condition, $a(t) \psi^2(t,x^k)$ coincides exactly with the background flat FLRW solution with $D_{(\alpha)} \to \bar{C}_{(\alpha)} D_{(\alpha)}$. Therefore, at a given time $t=t_0$, $\psi(t_0,x^k)$ is completely determined by $\{\bar{C}_{(\alpha)}(x^k)\}_{\alpha=1,\dots,N}$. 
In other words, if we require $\psi(t_{0},x^{k})$ to satisfy
\begin{equation}
    \psi(t_0,x^k) = f(x^k)
\end{equation}
for a given function $f(x^{k})$, 
this gives 
a single constraint for $\{\bar{C}_{(\alpha)}(x^k)\}_{\alpha=1,\dots,N}$.
In particular, if we require
\begin{equation}
\psi(t_0,x^k)=1,\label{pureentropy0re}
\end{equation}
we will obtain a constraint for $\{\bar{C}_{(\alpha)}(x^k)\}_{\alpha=1,\dots,N}$, which 
is independent of the adiabatic condition and can be regarded as an isocurvature condition.

\subsection{Leading-order solution with slicing conditions}
\subsubsection{Uniform e-folding slice}
The condition for the uniform e-folding slice ($\mathscr{N}$ slice) is given by
\begin{equation}
    \alpha K=-3\frac{\dot{a}}{a}.\label{Nslice}
\end{equation}
Taking this slice, we find the following relation from Eq.~(\ref{beyondlytho(1)hubble3}):
\begin{equation}
    \psi=\psi(x^k).\label{Npsi}
\end{equation}
Moreover, from Eq.~(\ref{beyondlytho(1)energy3}), we find that the energy densities for the fluid components are expressed by 
\begin{equation}
    {}^{(0)}\rho_{(\alpha)}(t,x^k)=c_{(\alpha)}(x^k)a^{-3\Gamma_{(\alpha)}}(t),\label{Nrho}
\end{equation}
where $c_{(\alpha)}(x^k)$ is an integral function for all fluid components.
Using this form of the energy densities, the mean curvature is formally expressed by
\begin{equation}
    {}^{(0)}K(t,x^k)=\sqrt{24\pi\sum_{\alpha=1}^{N}c_{(\alpha)}(x^k)a^{-3\Gamma_{(\alpha)}}(t)}.\label{NK}
\end{equation}
Combining Eqs.~(\ref{Nslice}) and (\ref{NK}), we can determine the lapse function under the $\mathscr{N}$ slice. We have also shown that the $\mathscr{N}$ slice is equivalent to neither the CMC slice nor the synchronous slice.

We define a density fluctuation for the component of the fluids as 
\begin{equation}
    {}^{(0)}\delta_{(\alpha)}(t,x^k)\equiv\frac{{}^{(0)}\rho_{(\alpha)}(t,x^k)-{}^{(b)}\rho_{(\alpha)}(t)}{{}^{(b)}\rho_{(\alpha)}(t)},\label{def:densityfluctuation}
\end{equation}
for all $\alpha$.
The minimum value of this function is $-1$.
The density fluctuation on $\mathscr{N}$ slice is reduced to
\begin{equation}
    {}^{(0)}\delta_{(\alpha)}=\bar{c}_{(\alpha)}(x^k)-1,\label{Ndelta}
\end{equation}
where we have assumed that the integral function is decomposed as follows:
\begin{equation}
    c_{(\alpha)}(x^k)=D_{(\alpha)}\bar{c}_{(\alpha)}(x^k).
\end{equation}
Eq.~(\ref{Ndelta}) represents the density fluctuation for the components of the fluids that are constant in time and independent of the function $\psi(x^k)$. 

\subsubsection{CMC slice}
Taking CMC slice, Eqs.~(\ref{beyondlytho(1)hubble3}) and (\ref{beyondlytho(1)energy3}) gives
\begin{equation}
    {}^{(0)}\rho_{(\alpha)}(t,x^k)=c^\prime_{(\alpha)}(x^k)\qty(a(t)\psi^2(t,x^k))^{-3\Gamma_{(\alpha)}},\label{CMCrho}
\end{equation}
where $c^\prime_{(\alpha)}(x^k)$ are integral function which comes from energy equation.
Putting Eq.~(\ref{CMCrho}) into Eq.~(\ref{beyondlytho(1)friedmann3}), Eq.~(\ref{beyondlytho(1)friedmann3}) is reduced to
\begin{equation}
    \qty(\frac{\dot{a}}{a})^2=\frac{8\pi}{3}\sum_{\alpha=1}^{N}c^\prime_{(\alpha)}(x^k)\qty(a(t)\psi^2(t,x^k))^{-3\Gamma_{(\alpha)}}.
\end{equation}
The left-hand side of the above equation corresponds to the left-hand side of the background Friedmann equation which is given by Eq.~(\ref{Haliltonianconstraint40}). Thus, we find the following relation:
\begin{equation}
\sum_{\alpha=1}^{N}D_{(\alpha)}a^{-3\Gamma_{(\alpha)}}(t)=\sum_{\alpha=1}^{N}c^\prime_{(\alpha)}(x^k)\qty(a(t)\psi^2(t,x^k))^{-3\Gamma_{(\alpha)}}.\label{CMCpsi}
\end{equation}
This equation is an algebraic equation for the function $\psi$.
Solving Eq.~(\ref{CMCpsi}) for $\psi$, we formally find the cosmological long wavelength solution:
\begin{equation}
    \psi(t,x^k)=\psi(a(t);D_{(\alpha)},\bar{c}^\prime_{(\alpha)}(x^k)),
\end{equation}
where we have decomposed the integral function:
\begin{equation}
    c^\prime_{(\alpha)}(x^k)=D_{(\alpha)}\bar{c}^\prime_{(\alpha)}(x^k).
\end{equation}
By adopting this solution to the energy density for the fluid components and incorporating the form of the energy density into the definition of the density fluctuation (\ref{def:densityfluctuation}), we observe that the density fluctuations for the fluid components exhibit both time and spatial dependence.
\subsubsection{Geodesic slice}
\label{2D}
As discussed in Sec.\ref{relationslice}, once we choose the geodesic slice, the proper time acts as the coordinate. Using the proper time as a time coordinate, the leading-order basic equations are reduced to
\begin{align}
  {}^{(0)}K^2=&24\pi {}^{(0)}E,\label{beyondlytho(1)friedmann4}\\
  {}^{(0)}K=&-3\frac{\PA_t\qty(a\psi^2)}{a\psi^2},\label{beyondlytho(1)hubble4}\\
  \PA_t{{}^{(0)}K}=&\qty[\frac{1}{3}{}^{(0)}K^2+4\pi{}^{(0)}\qty(E+S^i_i)],\label{beyondlytho(1)const4}\\
  \PA_t\qty(a^3\psi^6{}^{(0)}\rho_{(\alpha)})=&a^3\psi^6\qty(\Gamma_{(\alpha)}-1){}^{(0)}\rho_{(\alpha)}{}^{(0)}K.\label{beyondlytho(1)energy4}
\end{align}
We fix the shift vector in the following discussion as $\beta^i=0$.

Eqs.~(\ref{beyondlytho(1)hubble4}) and~(\ref{beyondlytho(1)energy4}) gives
\begin{equation}
  {}^{(0)}\rho_{(\alpha)}=C_{(\alpha)}(x^k)\qty(a\psi^2)^{-3\Gamma_{(\alpha)}}, \label{LWLsonfored1}
\end{equation} 
where $C_{(\alpha)}(x^k)$ is integral functions. 
By using Eqs.~(\ref{beyondlytho(1)friedmann4}) and~(\ref{LWLsonfored1}), the Hamiltonian constraint reduces to

\begin{equation}
  \qty[\frac{\PA_t\qty(a\psi^2)}{a\psi^2}]^2=\frac{8\pi}{3}\sum_{\alpha=1}^{N}C_{(\alpha)}(x^k)\qty(a\psi^2)^{-3\Gamma_{(\alpha)}}.
  \label{eq:Hconstraintperturbed0}
\end{equation}

In App.~\ref{App.3}, we have expanded the basic equations to the next-to-leading order, obeying the gradient expansion method. Expressing the integral form of the Euler equation at $\order{\epsilon}$, we have shown that this individual spatial component of a four-velocity $u^i$ under the geodesic slice and normal coordinates, i.e., Gaussian normal coordinates, has a decaying mode and the term which contains non-trivial time dependence at $\order{\epsilon}$. Since the pressure gradient contributes to the last term, it may play a crucial role in the dynamics of PBH formation.

\section{Long-wavelength solutions in two-fluid systems}
\label{sec4}
A two-fluid system is not only the simplest among the multi-fluid systems, but also provides physical insights into the behavior of the long-wavelength perturbations unique to 
multi-component systems.
Moreover, this practically gives a reasonable approximation to many physically realistic and cosmologically motivated 
multi-component systems. 
Here, we consider nonlinear long-wavelength solutions in two-fluid systems. The specific case of adiabatic and entropy perturbations of matter and radiation on super-horizon-scales in linear perturbation theory is discussed in Wands, Malik, Lyth, and Liddle~\cite{Wands:2000dp}.

\subsection{Background solutions}
The Friedmann equation for the background solution becomes
\begin{equation}
  \qty(\frac{\dot{a}}{a})^2=\frac{8\pi}{3}\qty[D_{(1)}a^{-3\Gamma_{(1)}}+D_{(2)}a^{-3\Gamma_{(2)}}].\label{Haliltonianconstraint4}
\end{equation}
The two-fluid system can be conveniently formulated by introducing the equality time $t_{\rm eq}$. 
Introducing the energy density and the scale factor at the equality time as
\begin{align}
  \frac{{}^{(b)}\rho_{\text{eq}}}{2}&\equiv D_{(1)}^{\Gamma_{(2)}/\qty(\Gamma_{(2)}-\Gamma_{(1)})}D_{(2)}^{\Gamma_{(1)}/\qty(\Gamma_{(1)}-\Gamma_{(2)})},\\
  a_{\text{eq}}&=\qty(\frac{D_{(1)}}{D_{(2)}})^{1/[3(\Gamma_{(1)}-\Gamma_{(2)})]},
\end{align}
the Friedmann equation~(\ref{Haliltonianconstraint4}) is reduced to
\begin{equation}
  \qty(\frac{\dot{a}}{a})^2=\frac{8\pi}{3}\frac{{}^{(b)}\rho_\text{eq}}{2}\qty[\qty(\frac{a}{a_\text{eq}})^{-3\Gamma_{(1)}}+\qty(\frac{a}{a_\text{eq}})^{-3\Gamma_{(2)}}],\label{anyfluidsbackgroundeq}
\end{equation}
in the background spacetime. 
Let us introduce a new variable $y$, which is defined by 
\begin{equation}
  y\equiv \frac{a}{a_\text{eq}}.\label{bgnewvariable}
\end{equation}
Then, Eq.~(\ref{anyfluidsbackgroundeq}) reduces to
\begin{equation}
  \frac{\dot{y}}{y}\frac{1}{\sqrt{y^{-3\Gamma_{(1)}}+y^{-3\Gamma_{(2)}}}}=\frac{H_\text{eq}}{\sqrt{2}},\label{anyfluidsbackgroundeq2}
\end{equation}
where $H_\text{eq}$ is the Hubble parameter at the equality time, which is defined by
\begin{equation}
  H_\text{eq}^2\equiv\frac{8\pi}{3}\rho_\text{eq}.
\end{equation}
The integral form of Eq.~(\ref{anyfluidsbackgroundeq2}) is given by
\begin{equation}
  \int_{0}^{y}\frac{\dd \tilde{y}}{\tilde{y}}
  \frac{\sqrt{2}}{\sqrt{\tilde{y}^{-3\Gamma_{(1)}}+\tilde{y}^{-3\Gamma_{(2)}}}}
   =H_\text{eq}\qty(t-t_\text{s}). \label{anyfluidsbackgroundeqsol}
\end{equation}
For convenience, we introduce
\begin{equation}
    q(y)\equiv \int_{0}^{y}\frac{\dd \tilde{y}}{\tilde{y}}
  \frac{\sqrt{2}}{\sqrt{\tilde{y}^{-3\Gamma_{(1)}}+\tilde{y}^{-3\Gamma_{(2)}}}}.\label{defq}
\end{equation}
Then, we can express the solution (\ref{anyfluidsbackgroundeqsol}) as follows:
\begin{equation}
  y=q^{-1}(H_\text{eq}\qty(t-t_\text{s})),
\end{equation}
and adopt this function as the background solution with a normalization condition
\begin{equation}
    a_\text{eq}=1.
\end{equation}
The normalization condition makes the integral constants equal
\begin{equation}
    D_{(1)}=D_{(2)}.
    \label{eq:D1=D2}
\end{equation}

\subsection{Long-wavelength solutions}
\subsubsection{Long-wavelength solutions with the geodesic slice}
For the perturbed system, if we choose the geodesic slice, the Hamiltonian constraint (\ref{eq:Hconstraintperturbed0}) reduces to
\begin{equation}
  \qty[\frac{\PA_t\qty(a\Psi^2)}{a\Psi^2}]^2=\frac{8\pi}{3}\qty[C_{(1)}(x^k)\qty(a\Psi^2)^{-3\Gamma_{(1)}}+C_{(2)}(x^k)\qty(a\Psi^2)^{-3\Gamma_{(2)}}].
\end{equation}
Introducing the following quantities,
\begin{align}
  {}^{(0)}\Xi(t,x^k)&\equiv a(t)\Psi^2(t,x^k),\\
  \frac{{}^{(0)}\rho_{\text{eq}}(x^k)}{2}&\equiv C_{(1)}(x^k)^{\Gamma_{(2)}/\qty(\Gamma_{(2)}-\Gamma_{(1)})}C_{(2)}(x^k)^{\Gamma_{(1)}/\qty(\Gamma_{(1)}-\Gamma_{(2)})},\\
  \tilde{H}_\text{eq}^2(x^k)&\equiv \frac{16\pi}3C_{(1)}(x^k)^{\Gamma_{(2)}/\qty(\Gamma_{(2)}-\Gamma_{(1)})}C_{(2)}(x^k)^{\Gamma_{(1)}/\qty(\Gamma_{(1)}-\Gamma_{(2)})},\\
  {}^{(0)}\Xi_{\text{eq}}(x^k)&\equiv \qty(\frac{C_{(1)}(x^k)}{C_{(2)}(x^k)})^{1/[3(\Gamma_{(1)}-\Gamma_{(2)})]},\label{HamiltonianconstleadingXI}
\end{align}
the leading-order Hamiltonian constraint takes the following form: 
\begin{equation}
    \qty[\frac{\PA_t{}^{(0)}\Xi}{{}^{(0)}\Xi}]^2=\frac{8\pi}{3}\frac{^{(0)}\rho_{\text{eq}}(x^k)}{2}\qty[\qty(\frac{{}^{(0)}\Xi}{{}^{(0)}\Xi_{\text{eq}}(x^k)})^{-3\Gamma_{(1)}}+\qty(\frac{{}^{(0)}\Xi}{{}^{(0)}\Xi_{\text{eq}}(x^k)})^{-3\Gamma_{(2)}}]
    .\label{finalhamiltonian}
\end{equation}
Eq.~(\ref{finalhamiltonian}) shows ${}^{(0)}\Xi(t,x^k)$ is monotonically increasing. 
This is integrated to give
\begin{equation}
  \int^{Y}_{0}\frac{\dd \tilde{Y}}{\tilde{Y}}\frac{\sqrt{2}}{\sqrt{\tilde{Y}^{-3\Gamma_{(1)}}+\tilde{Y}^{-3\Gamma_{(2)}}}}=\tilde{H}_\text{eq}(x^k)(t-{}^{(0)}\tilde{t}_\text{s}(x^k)),\label{LWLsolintegralorder1}
\end{equation}
where we have introduced
\begin{align}
  Y\equiv\frac{{}^{(0)}\Xi(t,x^k)}{{}^{(0)}\Xi_\text{eq}(x^k)}
\end{align}
and $\tilde{t}_\text{s}=\tilde{t}_\text{s}(x^k)$ is the initial time which satisfies $Y(\tilde{t}_\text{s}(x^k),x^k)=0$.
Eq.~(\ref{LWLsolintegralorder1}) can be rewritten in terms of $q$ as 
\begin{equation}
  Y\qty(t,x^k)=q^{-1}\qty(\tilde{H}_\text{eq}(x^k)\qty(t-{}^{(0)}\tilde{t}_\text{eq}(x^k))).\label{leadingorderlwlsol}
\end{equation}
By using this solution and the background solution, we find the leading-order conformal factor as follows:
\begin{equation}
    \Psi(t,x^k)=\sqrt{\frac{{q^{-1}\qty(\tilde{H}_\text{eq}(x^k)(t-{}^{(0)}\tilde{t}_\text{s}(x^k))){}^{(0)}\Xi_\text{eq}(x^k)}}{q^{-1}(H_{\text{eq}}(t-t_{\text{s}}))}}.
\end{equation}

Determining the long-wavelength solution for the conformal factor $\Psi(t,x^k)$ needs to fix the functions ${}^{(0)}\Xi_\text{eq}(x^k),~\tilde{H}_\text{eq}(x^k),~\text{and}~\tilde{t}_\text{s}(x^k)$. The parameter ${}^{(0)}\Xi_\text{eq}(x^k)$ is fixed by the ratio $C_{(1)}(x^k) / C_{(2)}(x^k)$, while $\tilde{H}_\text{eq}(x^k)$ is determined by their product. Accordingly, both $C_{(1)}(x^k)$ and $C_{(2)}(x^k)$ need to be uniquely specified.

\subsubsection{Asymptotic behaviours of the curvature and density perturbations}
\label{asymmptotic}
Let us consider the fluid (1) dominated limit of the function $q(Y)$.
The limit corresponds to the condition $Y\ll1$.

Denoting the integrand in the left-hand side of Eq.~(\ref{LWLsolintegralorder1}) by $F$, we have
\begin{equation}
    q(Y)=\int_0^YF(\tilde{Y})\dd \tilde{Y}.
\end{equation}
In the limit of $Y\ll1$, the function $F(Y)$ is expanded as
\begin{equation}
    F(Y)=\sqrt{2}Y^{\qty(3\Gamma_{(1)}/2)-1}+\dots.
\end{equation}
Therefore, we find the asymptotic behaviour of the function $q(Y)$ as
\begin{equation}
    q(Y)\simeq\frac{2\sqrt{2}}{3\Gamma_{(1)}}Y^{3\Gamma_{(1)}/2}.
\end{equation}
Thus, we find that, in the limit of $Y\rightarrow0$, the Hamiltonian constraint is expressed by
\begin{equation}
    \frac{2\sqrt{2}}{3\Gamma_{(1)}}Y^{\qty(3\Gamma_{(1)}/2)}\simeq\tilde{H}_\text{eq}(x^k)\qty(t-\tilde{t}_\text{s}(x^k)).
\end{equation}
The density contrast of the fluid (1), ${}^{(0)}\delta_{(1)}$, is expanded as
\begin{align}{}^{(0)}\delta_{(1)}
    &\simeq\frac{1}{\qty(t-\tilde{t}_\text{s}(x^k))^2}\qty[\qty(t-t_\text{s})^2-\qty(t-\tilde{t}_\text{s}(x^k))^2].
\end{align}
Assuming that we can neglect a decaying mode, the following relation is postulated 
\begin{equation}
    \tilde{t}_\text{s}(x^k)=t_\text{s}.
\end{equation}
We choose the origin of the time coordinate so that ${}^{(b)}H_\text{eq}t_\text{s}=0$.
Following the assumption, we find that, in the limit of $t\rightarrow t_\text{s}$, the zeroth order of the conformal factor as
\begin{align}
    \Psi&\rightarrow{}^{(0)}\Xi_\text{eq}^{1/2}(x^k)\qty(\frac{\tilde{H}_\text{eq}(x^k)}{H_\text{eq}})^{1/(3\Gamma_{(1)})}={}^{(0)}\Xi_\text{eq}^{1/2}(x^k)h^{1/\qty(3\Gamma_{(1)})}(x^k)=\exp\qty[-\frac{1}{2}\delta\mathscr{N}_{(1)}].\label{earlylimit}
\end{align}
From Eq.~(\ref{def:densityfluctuation}), the density fluctuation is reduced to
\begin{equation}
    {}^{(0)}\delta_{(\alpha)}\to\bar{C}_{(\alpha)}\Psi^{-6\Gamma_{(\alpha)}}(x^k)-1=\exp\qty[3\Gamma_{(\alpha)}\qty(\delta\mathscr{N}_{(1)}-\delta\mathscr{N}_{(\alpha)})]-1.\label{earlydelta}
\end{equation}
Therefore, $\Psi$ becomes constant in time in this limit.
In this limit, the contribution of fluid (1) dominates that of fluid (2), so that the curvature perturbation is governed by the perturbation of fluid (1). Hence, the system approaches the adiabatic regime.

Using these forms of the solutions with no decaying mode, the background and cosmological long wavelength solution for a two-fluid system are reduced to
\begin{align}
    a(t)&=q^{-1}(H_\text{eq}t),\\
    \Psi(t,x^k)&=\sqrt{\frac{q^{-1}(hH_\text{eq}t){}^{(0)}\Xi_\text{eq}(x^k)}{q^{-1}(H_\text{eq}t)}}.\label{Psifinal}
\end{align}

Next, we consider the late-time limit. In this limit, the function $q(Y)$ is approximated by
\begin{equation}
    q(Y)\simeq \frac{2\sqrt{2}}{3\Gamma_{(2)}}Y^{(3\Gamma_{(2)})/2}.
\end{equation}
Therefore, in the late time limit, the conformal factor $\Psi$ is reduced to
\begin{align}
    \Psi\rightarrow{}^{(0)}\Xi_\text{eq}^{1/2}(x^k)h^{1/(3\Gamma_{(2)})}(x^k)=\exp\qty[-\frac{1}{2}\delta\mathscr{N}_{(2)}],\label{latelimit}
\end{align}
which means $\Psi$ becomes constant in time. This arises from the fact that, in the late-time limit, fluid (2) becomes dominant over fluid (1), making this limit equivalent to the adiabatic one. The density fluctuation for fluid ($\alpha$) is then reduces to
\begin{equation}
    {}^{(0)}\delta_{(\alpha)}\to\bar{C}_{(\alpha)}\Psi^{-6\Gamma_{(\alpha)}}(x^k)-1=\exp\qty[3\Gamma_{(\alpha)}\qty(\delta\mathscr{N}_{(2)}-\delta\mathscr{N}_{(\alpha)})]-1,\label{latedelta2}
\end{equation}
at a late time. Therefore, the density fluctuation also becomes constant in time.

Therefore, both
$\Psi$ and ${}^{(0)}\delta_{(\alpha)}$ become time-independent in the early- and late-time limits, with their asymptotic behavior governed by the time evolution during the intermediate transition from early to late times.

\subsubsection{Curvature perturbations and density perturbations}
\label{two:dynamics}
In two-fluid systems, we find the leading-order solution is given by Eq.~(\ref{Psifinal}). Rewriting the solution by using the functions $h(x^k)$, $\delta\mathscr{N}_{(1)}$ and $\delta\mathscr{N}_{(2)}$, we obtain
\begin{align}
    \Psi&=\sqrt{\frac{q^{-1}(hH_\text{eq}t)}{q^{-1}(H_\text{eq}t)}}\exp\qty[-\frac{1}{2}\qty(\frac{\Gamma_{(1)}}{\Gamma_{(1)}-\Gamma_{(2)}}\delta\mathscr{N}_{(1)}-\frac{\Gamma_{(2)}}{\Gamma_{(1)}-\Gamma_{(2)}}\delta\mathscr{N}_{(2)})],\\
    &=\sqrt{\frac{q^{-1}(hH_\text{eq}t)}{q^{-1}(H_\text{eq}t)}}\exp\qty[-\frac{1}{2}\qty(\frac{\Gamma_{(1)}}{\Gamma_{(1)}-\Gamma_{(2)}}\delta\mathscr{N}^\text{nad}_{(1)}-\frac{\Gamma_{(2)}}{\Gamma_{(1)}-\Gamma_{(2)}}\delta\mathscr{N}^\text{nad}_{(2)}+\delta\mathscr{N}^\text{ad})]
\end{align}
where we choose the origin of the time coordinate as $H_\text{eq}t_\text{s}=0$. This relation shows that the curvature perturbation depends on both an adiabatic perturbation and an entropy perturbation, since the function $\delta\mathscr{N}_{(\alpha)}$  is expressed by
\begin{equation}
\delta\mathscr{N}_{(\alpha)}=\delta\mathscr{N}^\text{nad}_{(\alpha)}+\delta\mathscr{N}^\text{ad}.
\end{equation}
Whereas, we find the shape of the density perturbations for an individual fluid as follows in two-fluid systems on the geodesic slice:
\begin{align}
    {}^{(0)}\delta_{(\alpha)}&=\qty(\frac{q^{-1}(hH_\text{eq}t)}{q^{-1}(H_\text{eq}t)})^{-3\Gamma_{(\alpha)}}\exp\qty[\frac{3\Gamma_{(1)}\Gamma_{(2)}}{\Gamma_{(1)}-\Gamma_{(2)}}\qty(\delta\mathscr{N}_{(1)}-\delta\mathscr{N}_{(2)})]\notag\\
    &=\qty(\frac{q^{-1}(hH_\text{eq}t)}{q^{-1}(H_\text{eq}t)})^{-3\Gamma_{(\alpha)}}h^2(x^k).\label{densityperturb.intwofluid}
\end{align}
Eq.~(\ref{densityperturb.intwofluid}) therefore implies that, in two-fluid systems, the density perturbations of individual components depend only on $h$ or $\delta\mathscr{N}_{(12)}$.
This independence also holds for the total density perturbation
$\delta_{\rm tot}$ by its definition.
In contrast, the curvature perturbation $\Psi$ for purely entropic perturbation depends on the definition of an adiabatic perturbation.
As a result, the local e-folding number $\delta\mathscr{N}=\zeta$ is not enough to specify the physically relevant density perturbations in multi-fluid systems.
This originates from the intrinsic non-adiabaticity,
which introduces independent physical degrees of freedom not captured by a single local expansion history.

\subsection{Adiabatic and isocurvature conditions}
\label{sec3f}

\subsubsection{Adiabatic condition}
Assuming that the adiabatic condition Eq.~(\ref{pureadiabatic}) is satisfied at an equality time $t_0=\tilde{t}_\text{eq}(x^k)$ in two fluid systems, the condition is expressed as
\begin{equation}
    {}^{(0)}\rho_{(\alpha)}(\tilde{t}_\text{eq}(x^k),x^k)={}^{(b)}\rho_{(\alpha)}(\tilde{t}_\text{eq}(x^k)+\Delta t(x^k)),\label{twofluidadi}
\end{equation}
for $\alpha=1,~2$.
The energy densities of the two components are equal to each other at the equality time. Therefore, one can find that, from Eq.~(\ref{twofluidadi}), the energy densities of the two components of the corresponding background solution are also equal to each other. Thus, the following relations are satisfied:
\begin{gather}
    \tilde{t}_\text{eq}(x^k)+\Delta t(x^k)=t_\text{eq}\\
    {}^{(0)}{\rho}_\text{eq}(\tilde{t}_\text{eq}(x^k),x^k)={}^{(b)}\rho_\text{eq},
\end{gather}
where 
\begin{equation}
    {}^{(b)}\rho_\text{eq}=2{}^{(b)}\rho_{(1)}(t_\text{eq})=2{}^{(b)}\rho_{(2)}(t_\text{eq}),
\end{equation}
and
\begin{equation}
    {}^{(0)}\rho_\text{eq}(\tilde{t}_\text{eq}(x^k),x^k)\equiv2{}^{(0)}\rho_{(1)}(\tilde{t}_\text{eq}(x^k),x^k)=2{}^{(0)}\rho_{(2)}(\tilde{t}_\text{eq}(x^k),x^k).
\end{equation}
The Hamiltonian constraint Eq.~(\ref{finalhamiltonian}) at the equality time shows
\begin{equation}
    {}^{(0)}\tilde{H}_\text{eq}^2(x^k)=\frac{8\pi}{3}{}^{(0)}{\rho}_\text{eq}(\tilde{t}_\text{eq}(x^k),x^k)=\frac{8\pi}{3}{}^{(b)}{\rho}_\text{eq}={}^{(b)}H_\text{eq}^2.
\end{equation}
Namely, ${}^{(0)}\tilde{H}_\text{eq}(x^k)$ is constant throughout space and takes the same value as the background one. In this situation, Eq.~(\ref{Psifinal}) is reduced to
\begin{equation}
    \Psi=\qty(\frac{C_{(1)}(x^k)}{C_{(2)}(x^k)})^{1/\qty[6\qty(\Gamma_{(1)}-\Gamma_{(2)})]},
\end{equation}
which means that the leading-order conformal factor (curvature perturbation) remains constant in time.
In Sec.\ref{sec:defofentropy}, we decompose the integral functions $C_{(\alpha)}(x^k)$ which arise in Eq.~(\ref{decompositionofintegralfunc.}) as
\begin{equation}
    C_{(\alpha)}(x^k)=D_{(\alpha)}\bar{C}_{(\alpha)}(x^k).
\end{equation}
Then, the local Hubble parameter is expressed by
\begin{equation}
    {}^{(0)}\tilde{H}_\text{eq}(x^k)={}^{(b)}H_\text{eq}\sqrt{\bar{C}_{(1)}(x^k)^{\Gamma_{(2)}/\qty(\Gamma_{(2)}-\Gamma_{(1)})}\bar{C}_{(2)}(x^k)^{\Gamma_{(1)}/\qty(\Gamma_{(1)}-\Gamma_{(2)})}}= {}^{(b)}H_\text{eq}h(x^k).
\end{equation}
Since the pure adiabatic condition makes ${}^{(0)}\tilde{H}_\text{eq}(x^k)$ equal to the background one, the function $h(x^k)$ is unity. Therefore, we find the relation between $\bar{C}_{(1)}(x^k)$ and $\bar{C}_{(2)}(x^k)$ as follows:
\begin{equation}
    \bar{C}_{(1)}^{-\Gamma_{(2)}}(x^k)\bar{C}_{(2)}^{\Gamma_{(1)}}(x^k)=1.
\end{equation}
Considering that the leading-order energy densities, which are expressed by
\begin{equation}
    {}^{(0)}\rho_{(\alpha)}(t,x^k)=D_{(\alpha)}\bar{C}_{(\alpha)}(x^k)\qty(a\Psi^2)^{-3\Gamma_{(\alpha)}},
\end{equation}
the energy densities are reduced to
\begin{equation}
    {}^{(0)}\rho_{(1)}(t,x^k)=\qty(\frac{D_{(2)}}{D_{(1)}})^{1/\qty[3\qty(\Gamma_{(1)}-\Gamma_{(2)})]}{}^{(b)}\rho_{(1)}(t),
\end{equation}
\begin{equation}
    {}^{(0)}\rho_{(2)}(t,x^k)=\qty(\frac{D_{(1)}}{D_{(2)}})^{1/\qty[3\qty(\Gamma_{(2)}-\Gamma_{(1)})]}{}^{(b)}\rho_{(2)}(t).
\end{equation}
Using the normalization condition for the scale factor at the equality time 
or Eq.~(\ref{eq:D1=D2}),
we find
\begin{equation}
    {}^{(0)}\rho_{(\alpha)}(t,x^k)={}^{(b)}\rho_{(\alpha)}(t),
\end{equation}
for $\alpha=1,~2$. This condition means that the components of the leading-order density fluctuation vanish for an adiabatic perturbation. Note that the density fluctuations are generated at the second order of the long wavelength approximation. 

\subsubsection{Primeval isocurvature condition}
\label{initially isocurvature}

This condition corresponds to $\delta\mathscr{N}_{(1)}=0$ in two-fluids systems discussed in Sec.~\ref{iso:inflation}. Then, Eq.~(\ref{earlylimit}) with vanishing $\delta\mathscr{N}_{(1)}$ implies
\begin{equation}
    \Psi \rightarrow 1,
\end{equation}
in the limit of $t\to0$.
Therefore, in the early universe, the pure entropy condition constructed from the long-wavelength dynamics of inflation coincides with the standard isocurvature condition.  
However, this equivalence holds only when the early time limit provides a 
good approximation; once the dynamics deviates from this limit, the method 
given in Sec.~\ref{iso:inflation} and the initially isocurvature condition are no longer identical.

\subsubsection{Initially isocurvature condition}
\label{entropycondition}

Following the normalization condition, $a_\text{eq}=1$, the Friedmann equation at the equality time is reduced to
\begin{align}
    H_\text{eq}^2=\frac{8\pi}{3}\rho_\text{eq}
    =\frac{16\pi}{3}D_{(1)}^{\Gamma_{(2)}/\qty(\Gamma_{(2)}-\Gamma_{(1)})}D_{(2)}^{\Gamma_{(1)}/\qty(\Gamma_{(1)}-\Gamma_{(2)})}
    =\frac{16\pi}{3}D_{(1)}.\label{eq:Heq=16pi/3}
\end{align}
The integral functions $\bar{C}_{(\alpha)}(x^k)$ are rewritten by using ${}^{(0)}\tilde{H}_\text{eq}(x^k)~\text{and}~{}^{(0)}\Xi_\text{eq}(x^k)$ as follows:
\begin{align}
    D_{(\alpha)}\bar{C}_{(\alpha)}(x^k)=\frac{3}{16\pi}{}^{(b)}H^2_\text{eq}h^2(x^k){}^{(0)}\Xi_\text{eq}^{3\Gamma_{(\alpha)}}(x^k),\label{eq:C1alpha}
\end{align}
for $\alpha=1,~2$.
The pure entropy condition Eq.~(\ref{pureentropy0re}) is reduced to the following equation: 
\begin{equation}
    \Psi(t_0,x^k)=\sqrt{\frac{q^{-1}({}^{(b)}H_\text{eq}h(x^k)t_0){}^{(0)}\Xi_\text{eq}(x^k)}{q^{-1}({}^{(b)}H_\text{eq}t)}}=1,\label{pureentropytwofluid}
\end{equation}
in two-fluid systems. 
This is solved for ${}^{(0)}\Xi_\text{eq}(x^k)$ as
\begin{equation}
    {}^{(0)}\Xi_\text{eq}(x^k)=\frac{q^{-1}({}^{(b)}H_\text{eq}t_0)}{q^{-1}({}^{(b)}H_\text{eq}h(x^k)t_0)}.\label{eq:entropy_Xi_eq}
\end{equation}

To obtain $\bar{C}_{(1)}(x^k)$ and $\bar{C}_{(2)}(x^k)$ that satisfy the entropy condition, one may proceed as follows. 
First, choose the value of $H_{\mathrm{eq}}$ and the function
$h(x^k)$. Next substitute the chosen $H_{\mathrm{eq}}$ and $\tilde{H}_\text{eq}(x^k)=H_{\mathrm{eq}} h(x^{k})$
into the right-hand side of Eq.~(\ref{eq:entropy_Xi_eq}).
Then, Eq.~(\ref{eq:entropy_Xi_eq}) gives the function ${}^{(0)}\Xi_{\mathrm{eq}}(x^k)$.
Using this together with $h(x^k)$, 
one can then compute $\bar{C}_{(1)}(x^k)$ and $\bar{C}_{(2)}(x^k)$ from Eq.~(\ref{eq:C1alpha}), where $D_{1} = D_{2}$ can be determined from Eq.~(\ref{eq:Heq=16pi/3}).
Thus, this algorithm yields $\bar{C}_{(1)}(x^k)$ and $\bar{C}_{(2)}(x^k)$ corresponding to the entropy perturbation. 

\section{Demonstration: matter - radiation systems}
\label{sec5}
In this section, we construct solutions in the zeroth order of long-wavelength expansion for the system of 
matter and radiation starting from the entropy initial condition for the demonstration of the scheme developed here.
\subsection{Background and leading-order solutions}
\begin{figure}
    \centering
    \includegraphics[scale=1.0]{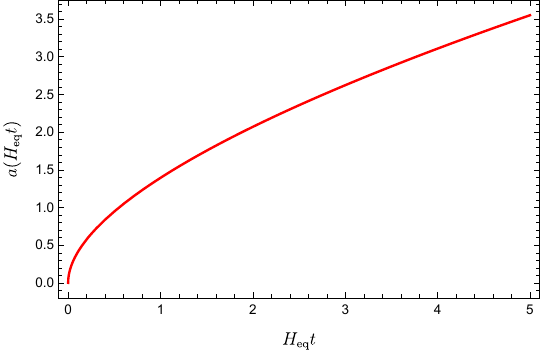}
    \caption{The evolution of scale factor $a(t)$ for matter and radiation fluid with  $a_\text{eq}=D_{(\mathrm{r})}/D_{(\mathrm{m})}=1,~H_\text{eq}t_\text{eq}=2\sqrt{2}\qty(2-\sqrt{2})/3$, radiation fluid with $\Gamma_{(\mathrm{r})}=4/3$, and matter fluid with $\Gamma_{(\mathrm{m})}=1$.}
    \label{fig:backgroundscalefactor}
\end{figure}

We consider a system filled with radiation ($\mathrm{r}$) and matter ($\mathrm{m}$). The EOS parameters in this system are 
\begin{equation}
  \Gamma_{(\mathrm{r})}=\frac{4}{3},~\Gamma_{(\mathrm{m})}=1.
\end{equation}
Under this system, we find the background solutions as follows:
\begin{equation}
  {}^{(\text{b})}\rho_{(\mathrm{r})}=D_{(\mathrm{r})}a^{-4},~{}^{(\text{b})}\rho_{(\mathrm{m})}=D_{(\mathrm{m})}a^{-3},
\end{equation} 
\begin{equation}
    q(y)=\frac{2}{3} \sqrt{2} \left(y\sqrt{y+1} -2 \sqrt{y+1}+2\right)=H_\text{eq}(t-t_\text{s}),
\end{equation}
where we have introduced $y$ as
\begin{equation}
    y=\frac{a}{a_\text{eq}},
\end{equation}
and $H_\text{eq},~\text{and}~a_\text{eq}$ are the Hubble parameter at equality time and the scale factor at an equality time, which gives 
\begin{align}
    a_\text{eq}=\frac{D_{(\mathrm{r})}}{D_{(\mathrm{m})}}.
\end{align}
The shape of this background solution is displayed in Fig. \ref{fig:backgroundscalefactor} with the condition $a_\text{eq}=1,~\text{and}~H_\text{eq}t_\text{eq}=2\sqrt{2}\qty(2-\sqrt{2})/3$.

The zeroth-order energy density for individual fluids is expressed as
\begin{equation}
  {}^{(0)}\rho_{(\mathrm{r})}=C_{(\mathrm{r})}(x^k){}^{(0)}\Xi^{-4},~{}^{(0)}\rho_{(\mathrm{m})}=C_{(\mathrm{m})}(x^k){}^{(0)}\Xi^{-3}.
\end{equation} 

\subsection{Time evolution of the curvature perturbation}

In Secs.~\ref{sec:defofentropy} and~\ref{isolation}, we have shown the definition of an entropy perturbation and three methods for isolating an adiabatic perturbation. Following each of these procedures, one can define a pure entropy perturbation associated with the corresponding method, and this can be imposed at an arbitrary initial time. In what follows, for each of the three approaches to isolate the adiabatic perturbation, we set the initial condition to be a pure entropy perturbation and investigate the subsequent time evolution of the relevant physical quantities. 

The integral functions are characterized by the function $h(x^k)$ in a two-fluid system and a pure entropy initial condition. To display the dynamics of the physical quantities with a pure entropy condition in the two-fluid system, we choose the function $h(x^k)$ and the other variables as
\begin{equation}
    h(r)=\sqrt{
  \frac{
    10^{\beta}
  }{
    \sqrt{
      2\pi \sigma_2^2
    }}
    \exp\!\left[
      -\frac{1}{2\sigma_2^2}
      (r - {}^{(b)}H_\text{eq}B)^2
    \right]
    + 1
  }\label{dimlessHubble}
\end{equation}
with $\beta=3$, $\sigma_{2}=3$ and ${}^{(b)}H_\text{eq}B=15$.
Also, the normalization condition for the background scale factor and the initial time is chosen as
\begin{gather}
   a_\text{eq}=\frac{D_{(\mathrm{r})}}{D_{(\mathrm{m})}}=1,~t_0=\frac{1}{1000}t_\text{eq}\qq{and} H_\text{eq}t_\text{eq}=\frac{2 \sqrt{2}}{3} (2 - \sqrt{2}),\label{seed2}
\end{gather}
respectively.

\begin{figure}
\begin{subfigure}{\linewidth}
    \centering
    \includegraphics[scale=1.0]{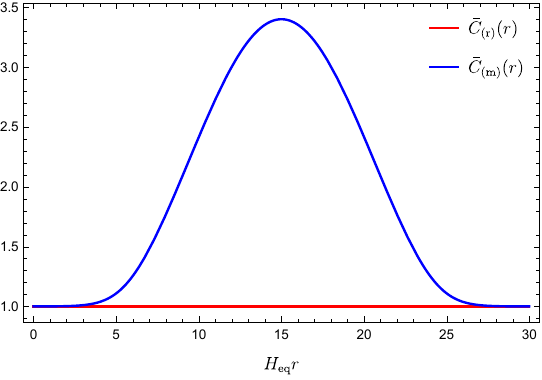}
    \caption{}
     \label{en:integrand}
\end{subfigure}
\begin{subfigure}{\linewidth}
    \centering
    \includegraphics[scale=1.0]{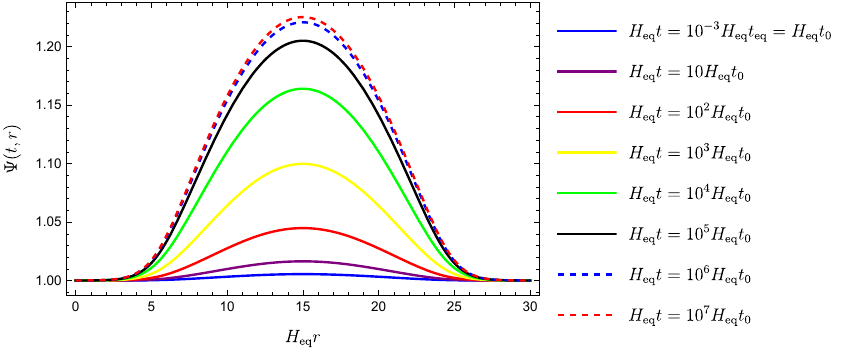}
    \caption{}
    \label{en:psi}
    \end{subfigure}
    \caption{The spatial dependence of the integral functions $\bar{C}_{(\mathrm{r})}$ and $\bar{C}_{(\mathrm{m})}$ (a), and the time evolution of the spherically symmetric curvature perturbation (b) satisfying the primeval isocurvature condition} 
    introduced in Sec.~\ref{iso:inflation} 
    under the conditions of Eqs.~(\ref{dimlessHubble}) and (\ref{seed2}) with the values $\beta=3,~\sigma_2=3,~{}^{(b)}H_\text{eq}B=15$. \label{en}
\end{figure}

First, we show the case of the primeval isocurvature condition discussed in Sec.~\ref{iso:inflation} in Figs.~\ref{en:integrand} and \ref{en:psi}.
Fig.~\ref{en:integrand} shows that the integral functions $\bar{C}_{(\mathrm{r})}$ and $\bar{C}_{(\mathrm{m})}$ obeying the algorithm discussed in Sec.~\ref{iso:inflation}. The function $\bar{C}_{(\mathrm{r})}$, which is a dominant component of the system in the early stage of the dynamics, is spatially constant, and its value is unity. Whereas, the function $\bar{C}_{(\mathrm{m})}$ has a spatial dependent profile which comes from the shape of $h(x^k)$.
Fig.~\ref{en:psi} shows the dynamics of the curvature perturbation obeying the algorithm discussed in Sec.~\ref{iso:inflation}. At the initial time, $\Psi(t,x^k)$ has a slight spatial dependence; therefore, at this initial time, the curvature perturbation is not isocurvature but slightly perturbed from unity. Note that the curvature perturbation is isocurvature only in the limit of $t\rightarrow0$ from Eq.~(\ref{earlylimit}). At late times, matter $(\mathrm{m})$ dominates radiation $(\mathrm{r})$, 
and the system asymptotically approaches an adiabatic state. 
Accordingly, the curvature perturbation becomes constant in time, 
reproducing the behavior expected for an adiabatic fluid. The late time behaviour of the curvature perturbation is determined by $\delta\mathscr{N}_{(2)}$ from Eq.~(\ref{latelimit}).

\begin{figure}
\begin{subfigure}{\linewidth}
    \centering
    \includegraphics[scale=1.0]{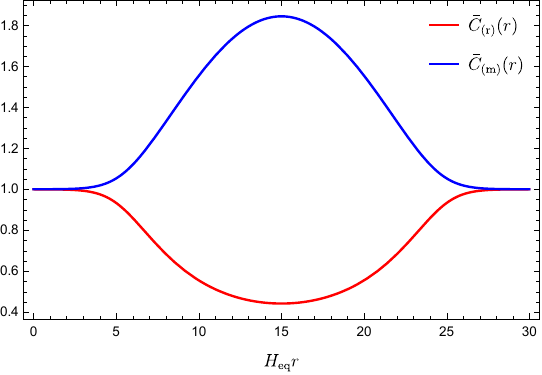}
     \caption{}
    \label{sym:integrand}
     \end{subfigure}
\begin{subfigure}{\linewidth}
    \centering
    \includegraphics[scale=1.0]{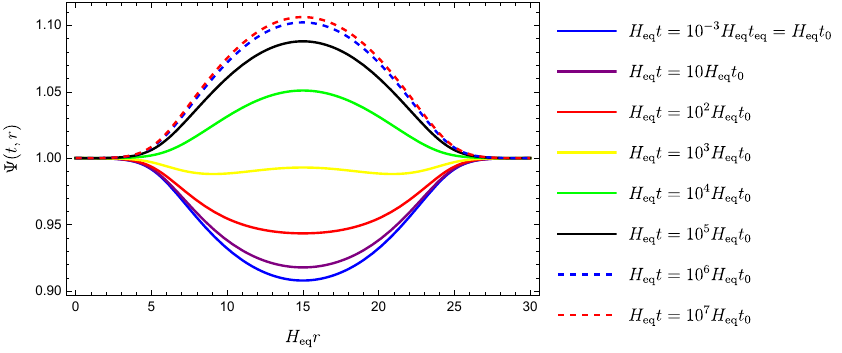}
   \caption{}
    \label{sym:psi}
    \end{subfigure}
    \caption{
    Same as Fig.~\ref{en} but for the zero-average condition.
    }
\end{figure}

Second, we show the case of the zero-average condition discussed in Sec.~\ref{sym:isolation} in Figs.~\ref{sym:integrand} and \ref{sym:psi}.
Fig.~\ref{sym:integrand} shows that the integral functions $\bar{C}_{(\mathrm{r})}$ and $\bar{C}_{(\mathrm{m})}$ obeying the algorithm discussed in Sec.~\ref{sym:isolation}. The functions $\bar{C}_{(\mathrm{r})}$ and $\bar{C}_{(\mathrm{r})}$ have functional forms whose convexities are opposite to each other. The shapes of the functions are determined by a profile of $h(x^k)$ from Eqs.~(\ref{sym:two1}) and~(\ref{sym:two2}) with $\delta\mathscr{N}^\text{ad}=0$.
Fig.~\ref{sym:psi} shows the dynamics of the curvature perturbation obeying the algorithm discussed in Sec.~\ref{sym:isolation}. At early times, $\Psi$ increases monotonically and exhibits a concave-downward behavior.
Near the equality epoch, $\Psi$ crosses $1$, the function becomes concave upward.
In the late-time limit, $\Psi$ asymptotically settles into a constant in time, as required by the effectively adiabatic perturbation condition. From Eqs.~(\ref{earlylimit}) and (\ref{latelimit}), we find that the early time and late time limits of the curvature perturbation $\Psi$ are reciprocal to each other which can be seen in Fig.~\ref{sym:psi}.

\begin{figure}
\begin{subfigure}{\linewidth}
    \centering
    \includegraphics[scale=1.0]{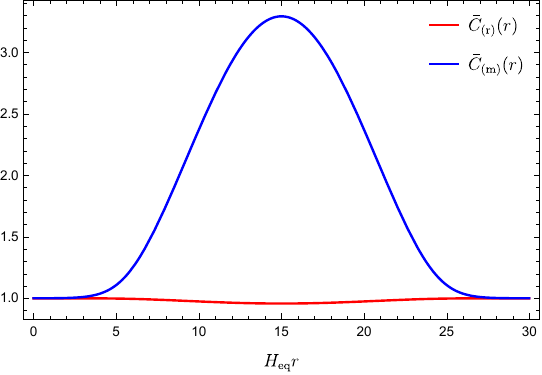}
    \caption{}
    \label{iso:integrand}
\end{subfigure}
\begin{subfigure}{\linewidth}
    \centering
    \includegraphics[scale=1.0]{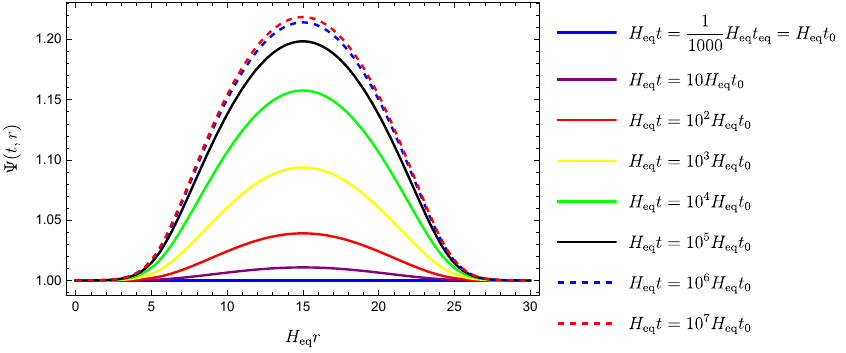}
    \caption{}
    \label{pureisopsi}
    \end{subfigure}
    \caption{
    Same as Fig.~\ref{en} but for the initially isocurvature condition.}
\end{figure}

Third, we show the case of the initially isocurvature condition discussed in Sec.~\ref{entropycondition} in Figs.~\ref{iso:integrand} and \ref{pureisopsi}.
Fig.~\ref{iso:integrand} shows the spatial dependence of the functions $\bar{C}_{(\mathrm{r})}(x^k)$ and $\bar{C}_{(\mathrm{m})}(x^k)$ in spherical symmetry following the algorithm which is discussed in Sec.~\ref{entropycondition}. The functional form of $h(x^k)$ is given by Eq.~(\ref{dimlessHubble}). The form of the function $\bar{C}_{(\mathrm{r})}$ is a spatial dependent form which is different from the case of Fig.~\ref{en:integrand}.
Fig.~\ref{pureisopsi} shows the curvature perturbation at leading order in terms of the gradient expansion following the procedure discussed in Secs.~\ref{iso:isolation} and \ref{entropycondition}. At the initial time where $H_\text{eq}t_0=H_\text{eq}t_\text{eq}/1000$, we impose the initial condition $\Psi = 1$, which makes the time dependence of $\Psi$ explicit as discussed in Sec.~\ref{entropycondition}. After that, it first grows slowly but later more rapidly with time evolution. In an adiabatic system, the curvature perturbation remains constant in time. 
However, in a two-fluid system where the total fluid is non-adiabatic, the curvature perturbation evolves even on super-horizon scales. 
At late times, matter $(\mathrm{m})$ dominates radiation $(\mathrm{r})$, 
and the system asymptotically approaches an adiabatic state. 
Accordingly, the curvature perturbation becomes constant in time, 
reproducing the behavior expected for the effectively adiabatic fluid.

\subsection{Time evolution of the density perturbations}
\begin{figure}
\begin{subfigure}{\linewidth}
    \centering
    \includegraphics[scale=1.0]{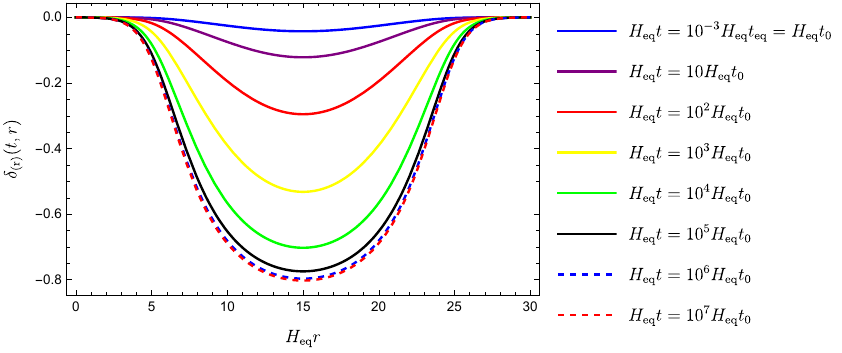}
    \caption{}
    \label{pureiso:delta1}
\end{subfigure}
\begin{subfigure}{\linewidth}
    \centering
    \includegraphics[scale=1.0]{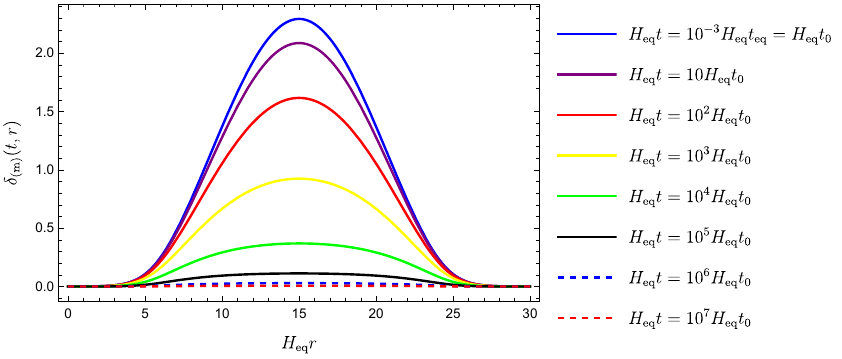}
    \caption{}
    \label{pureiso:delta2}
\end{subfigure}
\begin{subfigure}{\linewidth}
    \centering
    \includegraphics[scale=1.0]{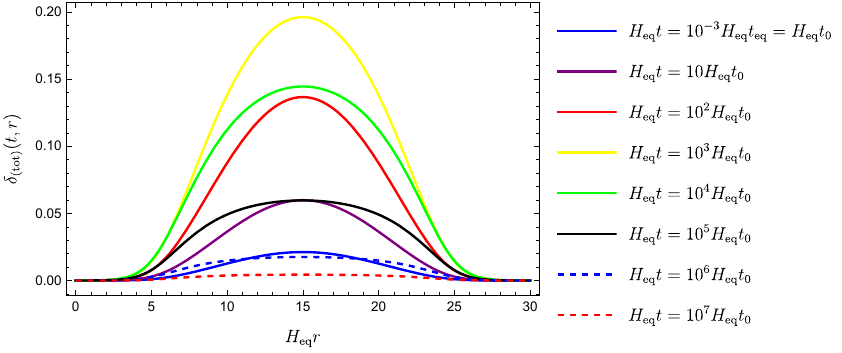}
    \caption{}
    \label{pureiso:deltatot}
    \end{subfigure}
    \caption{
    The snapshots of the time evolution of the spherically symmetric density perturbation for the radiation (r) (a), matter (m) (b), and total fluid (c) under the conditions of Eqs.~(\ref{dimlessHubble}) and (\ref{seed2}) with the values $\beta=3,~\sigma_2=3,~{}^{(b)}H_\text{eq}B=15$.}
\end{figure}

 As discussed in Sec.~\ref{two:dynamics}, although the dynamics of the curvature perturbation depend on how we isolate an adiabatic perturbation, the dynamics of the density perturbation do not. The temporal evolution of the energy density perturbations for each fluid component is presented in Figs.~\ref{pureiso:delta1}, \ref{pureiso:delta2}, and \ref{pureiso:deltatot}. Fig.~\ref{pureiso:delta1} illustrates that the density perturbation associated with radiation $(\mathrm{r})$ grows negatively with time. As we can see from Eq.~(\ref{earlydelta}), the density perturbation of radiation vanishes in the early time limit. At late times, when the universe becomes matter-dominated, the radiation perturbation ${}^{(0)}\delta_{(\text{r})}$ attains a time-independent negative profile.
Fig.~\ref{pureiso:delta2} shows the density perturbation of matter $(\mathrm{m})$. Initially, the density fluctuation is non-linearly large. As we discussed Eq.~(\ref{latedelta2}), the time evolution also exhibits a monotonic decrease and approaches zero at late times. 
Fig.~\ref{pureiso:deltatot} shows the total density fluctuation.
This function starts very small, increases to a maximum profile, decreases, and finally approaches zero at late times. Since the density fluctuation at $\order{\epsilon^0}$ vanishes in the adiabatic fluid \cite{Harada:2015yda}, this behavior recovers the adiabatic system. The time dependence of the fluctuation around the equality time shows a non-adiabaticity and affects the dynamics of the next-to-leading order in terms of the long-wavelength approximation $\epsilon$.

\section{Conclusion}
\label{sec6}
We have constructed cosmological nonlinear perturbation solutions for multi-fluid systems on super-horizon scales using the ADM formalism and the spatial gradient expansion method on the general coordinates.
Using these solutions, we have investigated the relations between different slicing conditions.
As a result, we find that the CMC, C, and UD slices are equivalent to $\order{\epsilon}$.
However, these slices are equivalent to neither the geodesic slice nor the $\mathscr{N}$ slice even at $\order{\epsilon^0}$.
By adopting the geodesic slice as the slicing condition, we fix the gauge completely and represent the cosmological long-wavelength solutions to $\order{\epsilon^0}$.
These growing-mode solutions are characterized by spatially dependent coefficients of the fluid energy densities, denoted by $\{C_{(\alpha)}(x^k)\}_{\alpha=1,\dots, N}$,
where $N$ is the number of the species.
The density perturbations can be nonlinearly large relative to the background flat FLRW spacetime, and thus can attain arbitrarily large amplitudes.
One of the most significant applications of these solutions is to construct initial data for the formation of PBHs, with or without imposing spherical symmetry.

Unlike the case of a single adiabatic fluid~\cite{Shibata:1999yda, Lyth:2004gb, Tanaka:2007gh, Harada:2015yda}, the long-wavelength solutions in this context are fixed by these $N$ functions $C_{(\alpha)}(x^k)$~rather than the curvature perturbation $\Psi$. By setting these functions $C_{(\alpha)}(x^k)$, we have shown that the zeroth-order solution can be expressed using the same function, $q$, as the background solution on the geodesic slice. As shown in Eq.~(\ref{LWLsonfored1}), in the present case, the energy density depends on time and spatial coordinates at the zeroth order. Consequently, from Eq.~(\ref{eq:Hconstraintperturbed0}), the curvature perturbation acquires time dependence from the zeroth order.

We have discussed the pure adiabatic and entropy conditions at a given time slice. Through the discussion, we define the adiabatic perturbation by Eq.~(\ref{pureadiabatic}) and the entropy perturbation by Eq.~(\ref{def:entropy}) in the multi-fluid system at any time slice. We have also presented three ways to separate an adiabatic perturbation and an entropy perturbation, which are given by Eqs.~(\ref{isolation:inflation}), (\ref{isolation:symmetric}), and (\ref{pureentropy0re}) in the multi-fluid system. In Sec.~\ref{sec4}, restricting our discussion to a two-fluid system, we derive long-wavelength solutions for two-fluid systems and have shown that in the geodesic slicing, the curvature perturbation depends on the way to isolate an adiabatic perturbation, but the density perturbation for each component $(\alpha)$ does not. 
Furthermore, in a matter–radiation system, we have demonstrated the time evolution of
the curvature perturbation, which satisfies the pure entropy condition during the radiation-dominated era in Sec.~\ref{sec5}. The curvature perturbation becomes constant in time in the limit of the early and late times, whereas depend on time around the equality time.
This time dependence originates from the non-adiabatic nature of the system.
We have also analyzed the time evolution of the density perturbations with the pure entropy conditions, showing that they become nonlinearly large and eventually approach the effectively adiabatic regime at late times.

The long-wavelength solutions discussed in this paper can be used as initial conditions for 
numerical simulations of PBH formation in multi-fluid systems. It will give us a deeper insight into the threshold of PBH
 formation by adopting these solutions to the conformal compactness function \cite{Harada:2024trx}. We leave these issues for future work.
 
\acknowledgments
We are grateful to Marco Bruni, Yuichiro Tada, Tsutomu Kobayashi, Shinji Mukohyama, Atsushi Naruko, Daisuke Yamauchi, Shuichiro Yokoyama, and Chul-Moon Yoo for fruitful discussions. 
This work was supported by Rikkyo University Special Fund for Research (H.I.). The work of T.H. was partially supported by JSPS KAKENHI Grants No. JP20H05853 and No. JP24K07027.

\appendix
\section{Linear limit of the nonlinear adiabatic condition}
\label{linearnonlinearentropy}
First, we review the entropy perturbation in the linear order.
The adiabatic condition in the cosmological context is given by
\begin{equation}
    p_\text{tot}=g(\rho_{\text{tot}}),
\end{equation}
which is a barotropic condition in the context of fluid dynamics. 
Irrespective of the gauge condition, the variation takes the following form
\begin{equation}
    \dv{p_\text{tot}}{\rho_\text{tot}}\delta\rho_\text{tot}=\dv{p_{(1)}}{\rho_{(1)}}\delta\rho_{(1)}+\dv{p_{(2)}}{\rho_{(2)}}\delta\rho_{(2)}\label{variation}
\end{equation}
in the two-fluid system.
Introducing the speed of sound as
\begin{align}
    {}^{(b)}c_{(\alpha)}^2&\equiv\dv{p_{(\alpha)}}{\rho_{(\alpha)}},\\
{}^{(b)}c_\text{tot}^2&\equiv\dv{p_\text{tot}}{\rho_\text{tot}}=\frac{{}^{(b)}c_{(1)}^2\dot{\rho}_{(1)}+{}^{(b)}c_{(2)}^2\dot{\rho}_{(2)}}{{}^{(b)}\dot{\rho}_{(1)}+{}^{(b)}\dot{\rho}_{(2)}},
\end{align}
Eq.~(\ref{variation}) becomes to
\begin{equation}
    \frac{{}^{(b)}c_{(1)}^2\dot{\rho}_{(1)}+{}^{(b)}c_{(2)}^2\dot{\rho}_{(2)}}{{}^{(b)}\dot{\rho}_{(1)}+{}^{(b)}\dot{\rho}_{(2)}}\qty(\delta\rho_{(1)}+\delta\rho_{(2)})={}^{(b)}c_{(1)}^2\delta\rho_{(1)}+{}^{(b)}c_{(2)}^2\delta\rho_{(2)}.
\end{equation}
This equation can be rearranged as follows
\begin{equation}
    \qty({}^{(b)}c_{(1)}^2-{}^{(b)}c_{(2)}^2)\qty(\frac{\delta\rho_{(1)}}{{}^{(b)}\dot{\rho}_{(1)}}-\frac{\delta\rho_{(2)}}{{}^{(b)}\dot{\rho}_{(2)}})=0.\label{adiabatic1}
\end{equation}
Assuming $c_{1}^2\ne c_{2}^2$, we find
\begin{equation}
    \frac{\delta\rho_{(1)}}{\dot{\rho}_{(1)}}-\frac{\delta\rho_{(2)}}{\dot{\rho}_{(2)}}=0.
\end{equation}
In the context of the linear perturbation theory, the entropy perturbation $S^{\text{L}}_{(\alpha\beta)}$ is defined by
\begin{equation}
    S^{\text{L}}_{(\alpha\beta)}\equiv-{}^{(b)}H\qty(\frac{\delta\rho_{(\alpha)}}{\dot{\rho}_{(\alpha)}}-\frac{\delta\rho_{(\beta)}}{\dot{\rho}_{(\beta)}}).\label{linearentropy}
\end{equation}
This perturbation at time $t=t_0$ is defined as adiabatic if
\begin{equation}
    S^{\text{L}}_{(\alpha\beta)}(t_0,x^k)=0
\end{equation}
holds for all pairs of components $\alpha$ and $\beta$.
This condition is equivalent to the following statement: A perturbation at time $t=t_0$ is called a \textit{linearized adiabatic perturbation}
if there exists a common spatially dependent time shift $\Delta t(t_0,x^k)$ such that,
for all components $\alpha$,
\begin{equation}
 \delta\rho_{(\alpha)}(t_0, x^k)
 = {}^{(b)}\dot{\rho}_{(\alpha)}(t_0)\,\Delta t(t_{0}, x^{k}).
 \label{linearized adiabatic perturbation}
\end{equation}

Next, we show that the nonlinear version of adiabatic and entropy perturbation recovers the linearised counterpart discussed above. 
Linearizing the pure adiabatic condition Eq.~(\ref{pureadiabatic}), the condition Eq.~(\ref{linearized adiabatic perturbation}) is recovered. Linearizing the definition of the entropy (isocurvature) perturbation $S_{(\alpha\beta)}$, Eq.~(\ref{entropyperturbation}), the approximated form of the condition is reduced to
\begin{equation}
    S_{(\alpha\beta)}\simeq \frac{\delta_{(\alpha)}}{3\Gamma_{(\alpha)}}-\frac{\delta_{(\beta)}}{3\Gamma_{(\beta)}}.
\end{equation}
This is the same as the conventional definition of the entropy perturbation, Eq.~(\ref{linearentropy}), in the context of the linear perturbation theory~\cite{Kodama:1984ziu, Kodama:1986fg, Kodama:1986ud}.

\section{Evolution equation for $\tilde{A}_{ij}$}
\label{leadingextrinsic}
The evolution equation for $\tilde{A}_{ij}$ at $\order{\epsilon}$ under the general slice is given by
\begin{equation}
\dot{\tilde{A}}_{ij}=-3\frac{{}^{(0)}\dot{\Xi}}{{}^{(0)}\Xi}\tilde{A}_{ij}.
\end{equation}
This equation gives an exact form of $\tilde{A}_{ij}$:
\begin{equation}
    \tilde{A}_{ij}=C_{ij}(x^k){}^{(0)}\Xi^{-3}.
\end{equation}
This solution is a decaying mode since we assume ${}^{(0)}\Xi$ is monotonically increasing.
Assuming that we can neglect the decaying mode, the order of  $\tilde{A}_{ij}$ is $\tilde{A}_{ij}=\order{\epsilon^2}$.

The leading order of the evolution equation for the induced metric is reduced to
\begin{equation}
    \dot{\tilde{\gamma}}_{ij}=-2{}^{(0)}\alpha\tilde{A}_{ij}=\order{\epsilon^2},
\end{equation}
under the general slice. This shows that following the assumption where the decaying mode of $\tilde{A}_{ij}$ is negligible, the time derivative of the induced metric at $\order{\epsilon}$ is also negligible.

\section{sub-leading-order expansion}
\label{App.3}
In leading order, we can neglect the spatial derivative terms in the Einstein equations and equations of motion for individual fluids. Since the Einstein equations and equations of motion are second-order partial differential equations, sub-leading-order equations include terms with second-order spatial derivatives. Considering sub-leading-order terms, we cannot neglect them. 

As shown in Appendix \ref{leadingextrinsic}, the leading order of $\tilde{A}_{ij}$ is a decaying mode under general slicing conditions and threading conditions. Following the assumption that we can neglect the decaying mode, we conclude that the leading order of the time derivative of the induced metric is also negligible. Therefore, the sub-leading order of the induced metric is of the second order: $\tilde{\gamma}_{ij}-\eta_{ij}=\order{\epsilon^2}$.
Based on the above assumption, we assume that basic variables are expanded as
\begin{align}
  \rho_{(\alpha)}&={}^{(b)}\rho_{(\alpha)}\qty(1+{}^{(0)}\delta_{(\alpha)})\qty(1+\delta_{(\alpha)})+\order{\epsilon^4},\\
p_{(\alpha)}&=\qty(\Gamma_{(\alpha)}-1)\rho_{(\alpha)}=\qty(\Gamma_{(\alpha)}-1){}^{(b)}\rho_{(\alpha)}\qty(1+{}^{(0)}\delta_{(\alpha)})\qty(1+\delta_{(\alpha)})+\order{\epsilon^4},\\
  v^i_{(\alpha)}&=\order{\epsilon},\\
  K&={}^{(0)}K\qty(1+\kappa)+\order{\epsilon^4},\\
  \tilde{A}_{ij}&=\order{\epsilon^2},\\
  \tilde{\gamma}_{ij}&=\eta_{ij}\qty(x^k)+h_{ij}+\order{\epsilon^4},\\
  \psi&=\Psi\qty(1+\xi)+\order{\epsilon^4},~\Xi=a\Psi^2\qty(1+2\xi)+\order{\epsilon^4},\\
  \alpha&={}^{(0)}\alpha+\chi+\order{\epsilon^4},\\
  \beta^i&=\order{\epsilon},
\end{align}
where
\begin{gather}
\delta_{(\alpha)}=\order{\epsilon^2},~\kappa=\order{\epsilon^2},~h_{ij}=\order{\epsilon^2},\\
  \Psi=\order{\epsilon^0},~\xi=\order{\epsilon^2},~\chi=\order{\epsilon^2},
\end{gather}
and
\begin{equation}
  {}^{(0)}\alpha\equiv1-\frac{\delta p(t,x^k)}{{}^{(0)}\rho_{\text{tot}}(t,x^k)+{}^{(0)}p_{\text{tot}}(t,x^k)},
\end{equation}
with general slices.

The ADM components of the energy-momentum tensor are expanded in the following form:
\begin{align}
E&={}^{(0)}\rho_{\text{tot}}+
\sum_{\alpha=1}^{N}{}^{(0)}\rho_{(\alpha)}\delta_{(\alpha)}+\order{\epsilon^4},\\
J_i&=\sum_{\alpha=1}^{N}\qty[\frac{1}{\alpha}\qty({}^{(0)}\rho_{(\alpha)}+{}^{(0)}p_{(\alpha)})\qty(v_{(\alpha)i}+\beta_i)]+\order{\epsilon^3},\\
S_{ij}&={}^{(0)}\Xi^2\sum_{\alpha=1}^{N}\qty(\Gamma_{(\alpha)}-1) {}^{(0)}\rho_{(\alpha)}\qty[\eta_{ij}\qty(1+2\zeta)+\delta_{(\alpha)}\eta_{ij}+h_{ij}]+\order{\epsilon^4},\\
S&=3\sum_{\alpha=1}^{N}\qty[\qty(\Gamma_{(\alpha)}-1){}^{(0)}\rho_{(\alpha)}\qty(1+\delta_{(\alpha)})]+\order{\epsilon^4}.
\end{align}

Then, the Hamiltonian constraint at sub-leading order is given by
\begin{equation}
\bar{\Delta}\Psi=-2\pi\Psi^5a^2\sum_{\alpha=1}^{N}\qty[{}^{(0)}\rho_{(\alpha)}\delta_{(\alpha)}-2\kappa{}^{(0)}\rho_{(\alpha)}]+\order{\epsilon^4}.
\end{equation} 
The momentum constraint at sub-leading order is given by
\begin{equation}
  \frac{2}{3}\tilde{D}_i{}^{(0)}K=-8\pi \sum_{\alpha=1}^{N}\qty[\frac{1}{\alpha}\Gamma_{(\alpha)}{}^{(0)}\rho_{(\alpha)}\qty(v_{(\alpha)i}+\beta_i)]+\order{\epsilon^3}.
\end{equation}
The evolution equation for the traceless part of the extrinsic curvature, Eq.~(\ref{cosmologicalconformaltracelessextrinsic}), reduces to
\begin{align}
  \dot{\tilde{A}}_{ij}+3\frac{{}^{(0)}\dot{\Xi}}{{}^{(0)}\Xi}\tilde{A}_{ij}&=\frac{1}{a^2\Psi^5}\biggl[{}^{(0)}\alpha\qty(\mathcal{R}_{ij}^\psi-\frac{a^2\Psi^5}{3}\mathcal{R}^\psi-\frac{8\pi}{a^2\Psi^4}S^\text{STF}_{ij})\notag\\&-\qty(\mathcal{D}_i\mathcal{D}_j{}^{(0)}\alpha-\frac{a^2\Psi^5}{3}\eta_{ij}\mathcal{D}_k\mathcal{D}^k{}^{(0)}\alpha)\biggr]+\order{\epsilon^4},
\end{align}
where we have used
\begin{equation}
  S_{ij}^\text{STF}\equiv S_{ij}-\frac{\gamma_{ij}}{3}S=\sum_{\alpha=1}^{N}\qty[\frac{1}{{}^{(0)}E+{}^{(0)}p_{(\alpha)}}\qty(J_{(\alpha)i}J_{(\alpha)j})^\text{STF}]+\order{\epsilon^4}.
\end{equation}
The evolution equation for the conformal factor, Eq.~(\ref{cosmologicalconformalconformalfactor}), reduces to
\begin{equation}
  6\dot{\xi}-3\frac{{}^{(0)}\dot{\Xi}}{{}^{(0)}\Xi}\qty(\frac{\chi}{{}^{(0)}\alpha}+\kappa)+\bar{\mathcal{D}}_k\beta^k=\order{\epsilon^4}.
\end{equation}
The evolution equation for the trace part of the extrinsic curvature, Eq.~(\ref{cosmologicalconformaltraceextrinsic}), reduces to
\begin{align}
  \frac{1}{{}^{(0)}K}\dot{\kappa}&=\frac{1}{3}{}^{(0)}\alpha\kappa+\frac{1}{3}\chi-\frac{\bar{\Delta}{}^{(0)}\alpha}{{}^{(0)}K^2}\notag\\
  &+\frac{{}^{(0)}\alpha}{6{}^{(0)}\rho_{\text{tot}}}
  \sum_{\alpha=1}^{N}\qty(3\Gamma_{(\alpha)}-2){}^{(0)}\rho_{(\alpha)}\qty(\chi+\delta_{(\alpha)}-\kappa)+\order{\epsilon^4}.
\end{align}
The evolution equation for the induced metric Eq.~(\ref{cosmologicalconformalthreemetevolution}) reduces to
\begin{equation}
\dot{h}_{ij}=-2\alpha_0\tilde{A}_{ij}+\eta_{kj}\bar{\mathcal{D}}_i\beta^k+\eta_{ik}\bar{\mathcal{D}}_j\beta^k-\frac{2}{3}\eta_{ij}\bar{\mathcal{D}}_k\beta^k+\order{\epsilon^4}.
\end{equation}

The energy equation for fluid (a) is  reduced to
\begin{align}
\PA_t&\qty[{}^{(0)}\Xi^3{}^{(0)}\rho_{(\alpha)}\qty{6\xi+\delta_{(\alpha)}-\frac{\Gamma_{(\alpha)}}{{}^{(0)}\alpha^2}\qty(\beta_k+v_{(\alpha)k})\qty(\beta^k+v_{(\alpha)}^k)}]+\frac{\Gamma_{(\alpha)}}{\sqrt{\eta}}\PA_k\qty[\sqrt{\eta}{}^{(0)}\Xi^3\rho_{(\alpha)0}\qty(\beta^k+v_{(\alpha)}^k)]\notag\\
=&-3{}^{(0)}\rho_{(\alpha)}{}^{(0)}\Xi^3\frac{{}^{(0)}\dot{\Xi}}{{}^{(0)}\Xi}\qty(1-\Gamma_{(\alpha)})\qty(\frac{\chi}{{}^{(0)}\alpha}+6\xi+\delta_{(\alpha)}+\kappa)-{}^{(0)}\Xi^3\frac{\PA_k{}^{(0)}\alpha}{{}^{(0)}\alpha}\Gamma_{(\alpha)}{}^{(0)}\rho_{(\alpha)}\qty(\beta^k+v_{(\alpha)}^k)\notag\\
&+\frac{{}^{(0)}\Xi^5}{3{}^{(0)}\alpha}\Gamma_{(\alpha)}{}^{(0)}\rho_{(\alpha)}{}^{(0)}K\eta_{l m}\qty(\beta^l+v_{(\alpha)}^l)\qty(\beta^m+v_{(\alpha)}^m)+\order{\epsilon^4}.
\end{align}
The Euler equation for fluid (a) is  reduced to
\begin{equation}
    \PA_t\qty[{}^{(0)}\Xi^3\Gamma_{(\alpha)}{}^{(0)}\rho_{(\alpha)}u_{(\alpha)j}]=-{}^{(0)}\alpha{}^{(0)}\Xi^3\qty[\qty(\Gamma_{(\alpha)}-1)\PA_j{}^{(0)}\rho_{(\alpha)}-\Gamma_{(\alpha)}{}^{(0)}\rho_{(\alpha)}\frac{\PA_j{}^{(0)}\alpha}{{}^{(0)}\alpha}]+\order{\epsilon^3}.
\end{equation}
The above equations are sub-leading-order basic equations under general gauge conditions.

To construct the sub-leading-order solutions, we adopt the geodesic slicing condition ($\alpha=1$) and normal coordinates ($\beta^i=0$) using the leading-order solution given by Eq.~(\ref{leadingorderlwlsol}).
Then, the Euler equation is reduced to
\begin{equation}
    \PA_t\qty[{}^{(0)}\Xi^3\Gamma_{(\alpha)}{}^{(0)}\rho_{(\alpha)}u_{(\alpha)j}]=-{}^{(0)}\Xi^3\qty(\Gamma_{(\alpha)}-1)\PA_j{}^{(0)}\rho_{(\alpha)}+\order{\epsilon^3}.\label{eulerfirstorder}
\end{equation}
The integral form of the Euler equation (\ref{eulerfirstorder}) under the geodesic slice and normal coordinates is given by
\begin{align}
    u_{(\alpha)j}=&-\frac{\Gamma_{(\alpha)}-1}{\Gamma_{(\alpha)}}{}^{(0)}\Xi^{-3\qty(1-\Gamma_{(\alpha)})}\int{}^{(0)}\Xi^{3\qty(1-\Gamma_{(\alpha)})}\frac{\PA_j\qty(C_{(\alpha)}(x^k)\Psi^{-6\Gamma_{(\alpha)}})}{C_{(\alpha)}(x^k)\Psi^{-6\Gamma_{(\alpha)}}}\dd t \notag\\
    &+{}^{(0)}\Xi^{-3\qty(1-\Gamma_{(\alpha)})}C_{(\alpha)j}(x^k)+\order{\epsilon^3},\label{fourvelocitylow}
\end{align}
where $C_{(\alpha)j}(x^k)$ is the integral function for individual fluids. The first term on the right-hand side of the above equation represents the pressure gradient. This term might contribute to the PBH formation. 
The upper script of $u_{(\alpha)j}$ is represented by
\begin{align}
    u^i_{(\alpha)}=
    &\eta^{ij}\biggl[\frac{1-\Gamma_{(\alpha)}}{\Gamma_{(\alpha)}}
    {}^{(0)}\Xi^{3\Gamma_{(\alpha)}-5}
    \int{}^{(0)}\Xi^{3\qty(1-\Gamma_{(\alpha)})}
    \frac{\PA_j\qty(C_{(\alpha)}(x^k)\Psi^{-6\Gamma_{(\alpha)}})}{C_{(\alpha)}(x^k)\Psi^{-6\Gamma_{(\alpha)}}}\dd t\notag\\
   &+{}^{(0)}\Xi^{3\Gamma_{(\alpha)}-5}C_{(\alpha)j}\qty(x^k)\biggr]+\order{\epsilon^3}.\label{fourvelocityup}
\end{align}
It is concluded that, since ${}^{(0)}\Xi$ is monotonically increasing, the term proportional to the integration function with the EOS parameter smaller than $\Gamma_{(\alpha)}<3/5$ is the decaying term, 
while the term representing the pressure gradient has a non-trivial time dependence.

Then, the sub-leading-order basic equations are reduced to the following form:
\begin{align}
&\bar{\Delta}\Psi=-2\pi\Psi^5a^2\sum_{\alpha=1}^{N}\qty[{}^{(0)}\rho_{(\alpha)}\delta_{(\alpha)}-2\kappa{}^{(0)}\rho_{(\alpha)}]+\order{\epsilon^4},\label{Hamiltonianconstgeonormal2nd}\\
 &\frac{2}{3}\bar{D}_i{}^{(0)}K=-8\pi \sum_{\alpha=1}^{N}\qty[\Gamma_{(\alpha)}{}^{(0)}\rho_{(\alpha)}v_{(\alpha)i}]+\order{\epsilon^3}\\
 &\dot{\tilde{A}}_{ij}+3\frac{{}^{(0)}\dot{\Xi}}{{}^{(0)}\Xi}\tilde{A}_{ij}=\frac{1}{a^2\Psi^5}\qty[{}^{(2)}R^\psi_{ij}-\frac{a^2\Psi^5}{3}{}^{(2)}R^\psi-\frac{8\pi}{a^2\Psi^4} {}^{(2)}S_{ij}]+\order{\epsilon^4},\\
 &6\dot{\xi}-3\frac{{}^{(0)}\dot{\Xi}}{{}^{(0)}\Xi}\kappa=\order{\epsilon^4},\\
 &\frac{\dot{\kappa}}{{}^{(0)}K}=\frac{1}{3}\kappa+\frac{1}{{}^{(0)}\rho_{\text{tot}}}\sum_{\alpha=1}^{N}\qty(3\Gamma_{(\alpha)}-2){}^{(0)}\rho_{(\alpha)}\qty(\delta_{(\alpha)}-\kappa)+\order{\epsilon^4},\\
 &\dot{h}_{ij}=-2\tilde{A}_{ij}+\order{\epsilon^4},\\
 &\PA_t\qty[{}^{(0)}\Xi^3{}^{(0)}\rho_{(\alpha)}\qty(6\xi+\delta_{(\alpha)})]
 =-{}^{(0)}\Xi^3{}^{(0)}\rho_{(\alpha)}\frac{{}^{(0)}\dot{\Xi}}{{}^{(0)}\Xi}\qty[3\qty(1-\Gamma_{(\alpha)})\qty(6\xi+\delta_{(\alpha)}+\kappa)]+\order{\epsilon^4},\\
 &\PA_t\qty[{}^{(0)}\Xi^3\Gamma_{(\alpha)}{}^{(0)}\rho_{(\alpha)}u_{(\alpha)j}]=-{}^{(0)}\Xi^3\qty(\Gamma_{(\alpha)}-1)\PA_j{}^{(0)}\rho_{(\alpha)}+\order{\epsilon^3}\label{eulerfirstorder2},
\end{align}
where we have used the following relation:
\begin{equation}
  u^t=1+\order{\epsilon^2},
\end{equation}
under geodesic slicing conditions and 
\begin{align}
    ^{(2)}R^\psi_{ij}&=-\frac{2}{\Psi}\bar{\mathcal{D}}_i\bar{\mathcal{D}}_j\Psi-\frac{2}{\Psi}\eta_{ij}\bar{\Delta}\Psi+\frac{6}{\Psi^2}\bar{\mathcal{D}}_i\Psi\bar{\mathcal{D}}_j\Psi-\frac{2}{\Psi^2}\eta_{ij}\bar{\mathcal{D}}_k\Psi\bar{\mathcal{D}}^k\Psi,\\
    ^{(2)}R^\psi&=-\frac{8}{\Psi^5 a^2}\bar{\Delta}\Psi,\\
    ^{(2)}S_{ij}&=\sum_{\alpha=1}^{N}\qty[\frac{1}{{}^{(0)}E_{(\alpha)}+{}^{(0)}p_{(\alpha)}}{}^{(2)}\qty(J_{(\alpha)i}J_{(\alpha)j})],\\
    ^{(2)}\qty(J_{(\alpha)i}J_{(\alpha)j})&=\qty(\Gamma_{(\alpha)}{}^{(0)}\rho_{(\alpha)})^2v_{(\alpha)i}v_{(\alpha)j},
\end{align}
under the normal coordinates. Solving Eqs.~(\ref{Hamiltonianconstgeonormal2nd})~-~(\ref{eulerfirstorder2}) yields a second-order long-wavelength solution. 
Although it is difficult, in general, to solve them analytically with arbitrary EOS parameters $\Gamma_{(\alpha)}$, they can be solved numerically by putting the integral form of the background solution and the zeroth-order solution into Eqs.~(\ref{Hamiltonianconstgeonormal2nd})~-~(\ref{eulerfirstorder2}).

\bibliographystyle{JHEP}
\bibliography{refs}

\end{document}